\documentstyle[12pt,epsfig]{article}
 1
\def\citenum#1{{\def\@cite##1##2{##1}\cite{#1}}}
\def\citea#1{\@cite{#1}{}}

\def\b{\beta}
\def\a{\alpha}

\def\D{\Delta}

\def\G{\Gamma}

\def\l{\lambda}

\def\O{\Omega}

\def\ra{\rightarrow}

\def\s{\sigma}

\def\({\left(}
\def\){\right)}

\def\citenum#1{{\def\@cite##1##2{##1}\cite{#1}}}
\def\citea#1{\@cite{#1}{}}

\def\l1vt{\vec{l_{1\perp}}}

\def\bt{b_{\perp}}

\def\bt2{$b^2_t$}

\def\jol1{$J_0(\,l_{1\perp}\,r_{\perp}\,)$}

\def\citea#1{\@cite{#1}{}}

\def\beq{\begin{equation}}
\def\eeq{\end{equation}}
\def\bea{\begin{eqnarray}}
\def\eea{\end{eqnarray}}

\def\eq#1{{Eq.~(\ref{#1})}}

\def\bbbz{{\mathchoice {\hbox{$\sf\textstyle Z\kern-0.4em Z$}}
{\hbox{$\sf\textstyle Z\kern-0.4em Z$}}
{\hbox{$\sf\scriptstyle Z\kern-0.3em Z$}}
{\hbox{$\sf\scriptscriptstyle Z\kern-0.2em Z$}}}}
\relax
\begin{document}
\newcounter{savefig}
\newcommand{\alphfig}{\addtocounter{figure}{1}%
\setcounter{savefig}{\value{figure}}%
\setcounter{figure}{0}%
\renewcommand{\thefigure}{\mbox{\arabic{savefig}-\alph{figure}}}}
\newcommand{\resetfig}{\setcounter{figure}{\value{savefig}}%
\renewcommand{\thefigure}{\arabic{figure}}}

\begin{titlepage}
\noindent
\begin{flushright}
TAUP 2561 - 99\\
  \today \\
\end{flushright}
\vspace{1cm}
\begin{center}
{\Large \bf {    THE SURVIVAL PROBABILITY}}\\[1.5ex]
{\Large \bf{ OF
  LARGE RAPIDITY GAPS IN}}\\[1.5ex]

{ \Large \bf {  A  THREE CHANNEL MODEL }}\\[6ex]

{\large E. G O T S M A N${}^{1)}$, E. L E V I N${}^{2)}$\,\,
and U.\,\,M A O R${}^{3)}$}
 \footnotetext{$^{1)}$ Email: gotsman@post.tau.ac.il .}
\footnotetext{$^{2)}$ Email: leving@post.tau.ac.il.}
\footnotetext{$^{3)}$ Email: maor@post.tau.ac.il.}\\[5.5ex]
{\it  School of Physics and Astronomy}\\
{\it Raymond and Beverly Sackler Faculty of Exact Science}\\
{\it Tel Aviv University, Tel Aviv, 69978, ISRAEL}\\[3.5ex]

\end{center}
~\,\,\,
\vspace{2cm}

{\large \bf Abstract:}

\par The values and energy dependence for the survival probability
$< \mid S \mid^2 >$  of
large rapidity
gaps (LRG)   are  calculated in a  three
channel model. This model includes   single and double
diffractive production, 
as well
as elastic rescattering. It is shown that
   $< \mid S \mid^2 >$ decreases with
increasing energy, in line with recent results for LRG dijet production at
the  Tevatron. This is  in spite of 
the weak  dependence on energy  of the
ratio $ (
\sigma_{el}
+
\sigma_{SD})/\sigma_{tot}$.

\end{titlepage}

\section{Introduction}

A  large rapidity gap ( LRG ) process is defined as one where  no
hadrons are produced in a sufficiently large rapidity region.
 Historically, both   Dokshitzer et al.\cite{Dok} and  Bjorken\cite{Bj},
 suggested utilizing LRG as a signature for Higgs
production in a  W-W fusion process, in hadron-hadron collisions.
It turns out that the LRG processes give a unique opportunity to
measure the high energy asymptotic behaviour  of the amplitudes  at short
distances,  where one  can use the
developed methods of perturbative QCD ( pQCD ) to calculate the
amplitudes.
Consider  a  typical LRG process - the production of two jets
with large  transverse momenta $ \vec{p}_{t1}\,\,\approx\,\,-
\vec{p}_{t2}\,\,\gg\,\,\mu $,  with a  LRG between the two jets.
$\mu$
is a typical mass scale of ``soft" interactions. We have
the reaction
\bea \label{I1}
& p (1)\,\,+\,\,p(2)\,\,\longrightarrow\,\,
M_1[ hadrons\,\,+\,\,jet_1(y_1,p_{t1})]&\\&
+\,\,LRG[ \Delta y = | y_1 - y_2|]
\,\,+\,\,M_2[ hadrons\,\,+\,\,jet_2(y_2,p_{t2})]&\nonumber
\eea
where $y_1$ and $y_2$ are the  rapidities  of the  jets and    $\Delta
y = | y_1 -
y_2 |\,\,\gg\,\,1$.  The production of two jets with LRG between them
can occur because:
\begin{enumerate}

\item  A  fluctuation in the rapidity distribution  of a  typical
inelastic event.
However, the probability for  such a fluctuation is proportional to
$ e^{ - \frac{\Delta y}{L}}$, where $L$ denotes  the value of the
correlation length.
We can evaluate $L\,\approx\,\frac{1}{\frac{d n}{d y}}$, where
$\frac{d n}{d y} $ is the number of particles per unit in
rapidity.   A  LRG means that $\Delta y \,\gg\,L$ and so the probability
is  small;

\item The exchange of a  colourless  state in QCD.
This exchange is  given by the amplitude of the high
energy interaction at short distances. We denote it as a ``hard" Pomeron
 in Fig.1.      We denote by  $F_s$ the ratio of the
cross section due to this  Pomeron exchange, to the typical inelastic
event  cross section
generated by  gluon exchange ( see Fig. 1 ).
In QCD we do not expect  this ratio   to decrease  as a function of the
rapidity gap $\Delta y = y_1 - y_2 $. For a BFKL Pomeron\cite{BFKL},
we expect an increase once  $\Delta y \,\gg\,\,1$.  Using
a  simple QCD model for the Pomeron , namely, that it can be approximated
by two gluon exchange\cite{LN},   Bjorken\cite{Bj} gave the
first estimate for $F_s \,\approx\,\,0.15 $, which is unexpectantly large.

\end{enumerate}

\begin{figure}
\centerline{\epsfig{file=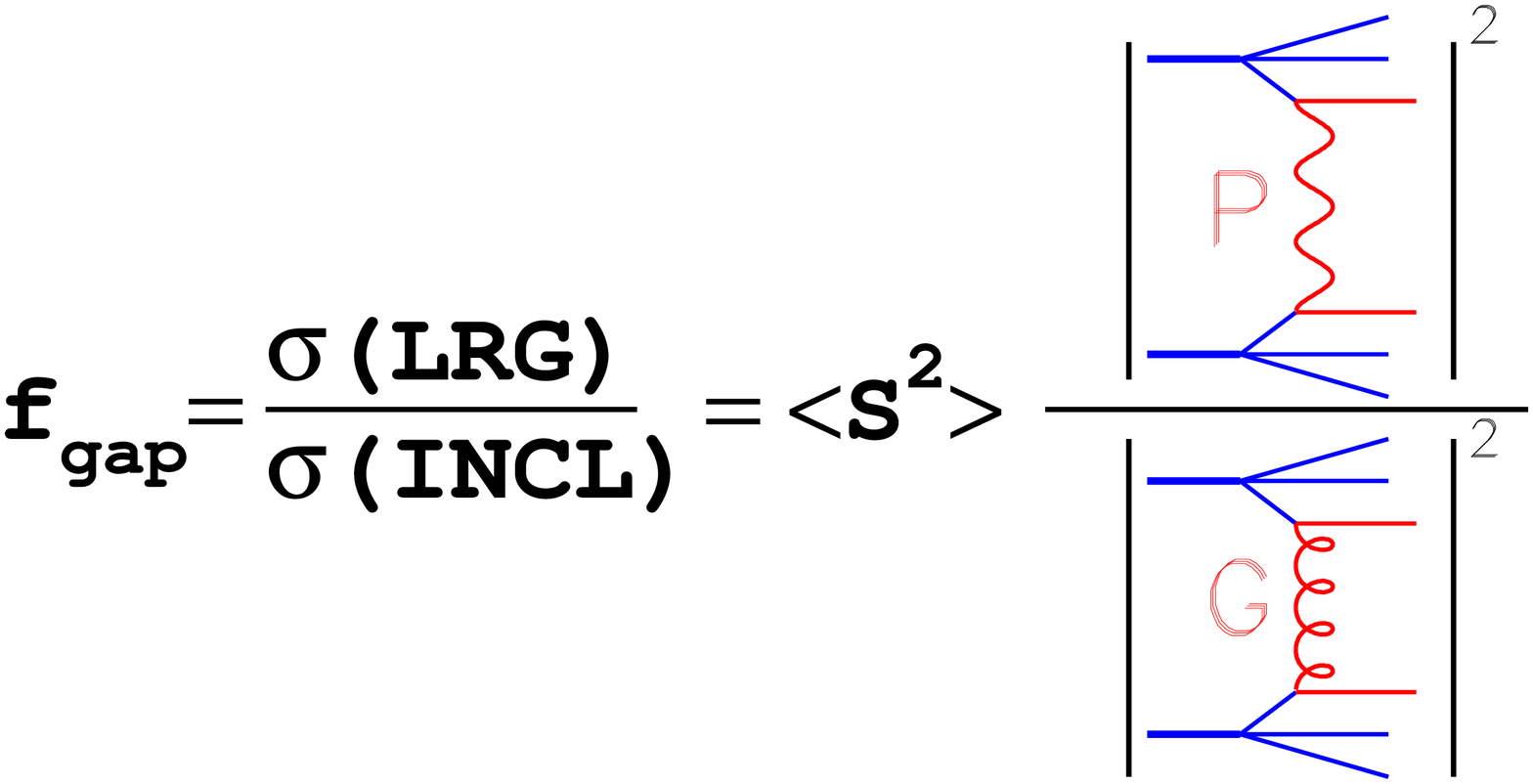,width=140mm}}
\caption{\it Pictorial definition of $f_{gap}$, where P and G represent
respectively, the exchange of a colour singlet and a colour octet. }
\label{Fig.1}
\end{figure}

As noted by Bjorken, we are not able to measure $F_s$ directly
in
a LRG
experiment. The experimentally measured  ratio of the number of
events with a LRG, to
the number of events without a LRG ( see Fig. 1 )
 is not equal to $F_s$,  but has to be modified  by an  extra factor
which we call the survival probability of LRG.
\beq \label{I2}
f_{gap} = < \mid S \mid^2 > \,\times\, F_{s}\,\,.
\eeq
The appearance of $ < \mid S \mid^2 > $ in \eq{I2} has a very
simple physical interpretation. It is the  probability that the
rapidity gap due to  Pomeron exchange, will not be filled by the
produced particles ( partons and/or hadrons ) from the rescattering
of  spectator partons ( see Fig.2a ), or from the emission of
bremsstrahlung gluons from partons taking part in the ``hard"
interaction, or from the ``hard" Pomeron ( see Fig. 2b ).
\beq \label{I3}
 < \mid S \mid^2 >\,\,\,=\,\,\, < \mid S_{bremsstrahlung} ( \,\Delta
y\,\,=\,\,|\,y_1\,\,-\,\,y_2\,|\,) \mid^2>\,\,\times\,\,
 < \mid S_{spectator}(\,s\,) \mid^2 >\,\,,
\eeq
where $s$ denotes  the total c.m. energy squared.

\begin{figure}
\begin{tabular}{c c}
\psfig{file= 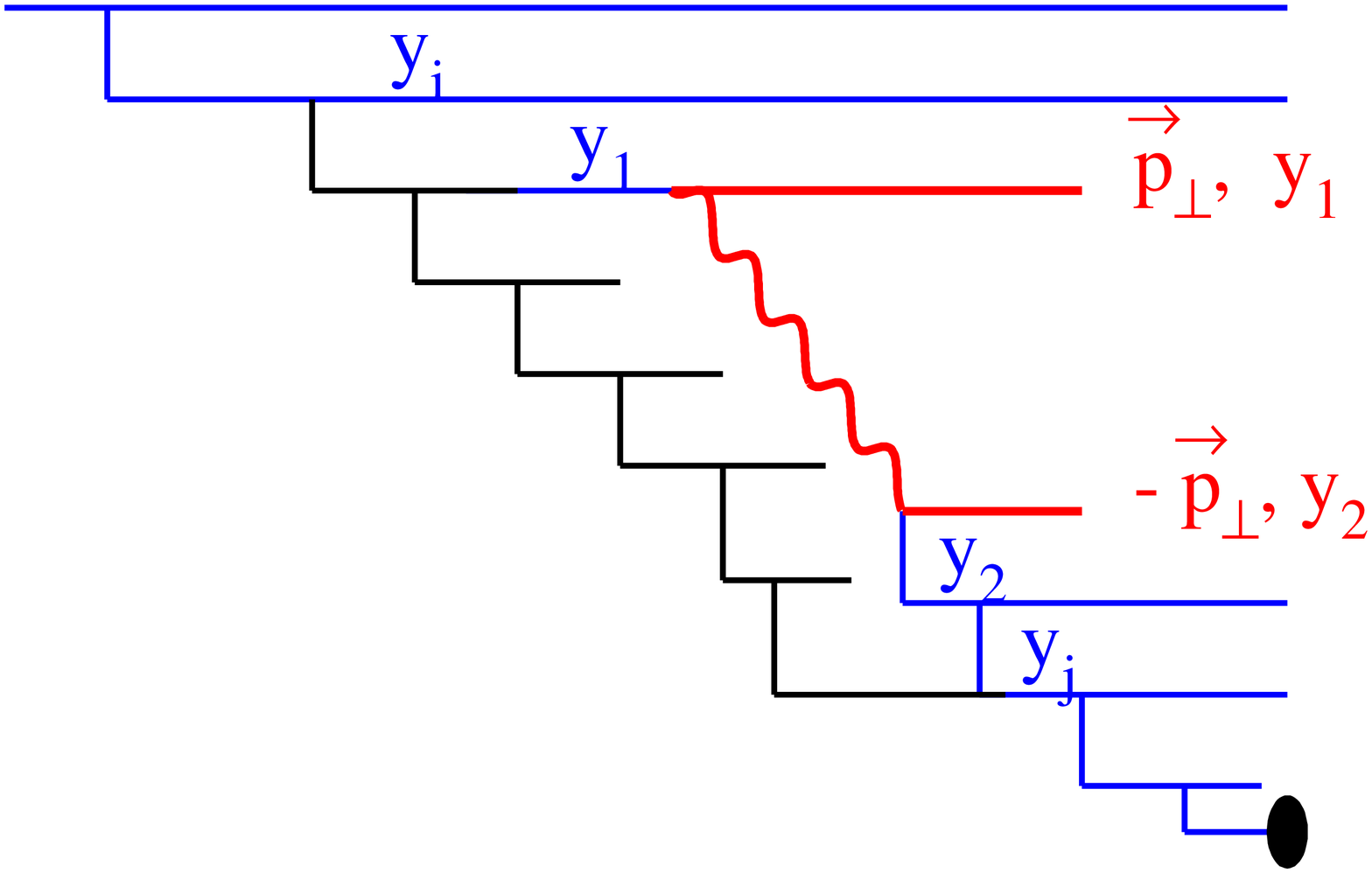, width=80mm,height=70mm} &\psfig{file=
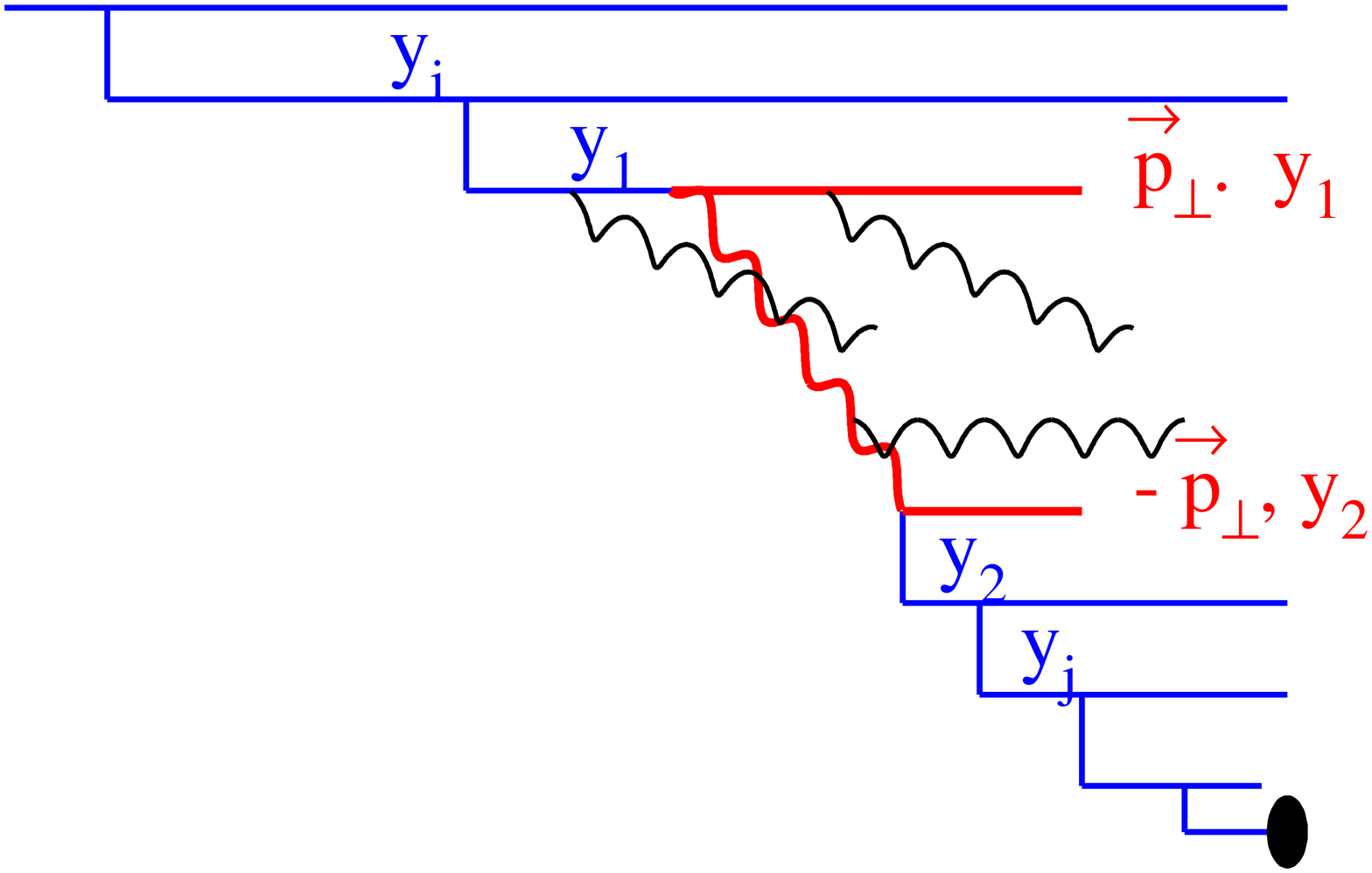,width=80mm,height=70mm}\\
Fig. 2-a & Fig. 2-b\\
\end{tabular}
\vspace{2cm}
\caption{ \it  Rescatterings of the  spectator partons ( Fig.2a ) and
emission of the bremsstrahlung gluons ( Fig.2b ) that fill the LRG .}
\label{fig2 }
\end{figure}

\begin{itemize}
\item\,\,\, $ < \mid S_{bremsstrahlung} ( \,\Delta
y\,) \mid^2>$ can be calculated in pQCD\cite{BREM}, it
depends on the kinematics of each specific process, and on the value of
the
LRG .

\item\,\,\,To calculate $< \mid S_{spectator}(\,s\,) \mid^2 >$ we need to
find the probability that all partons with rapidity $y_i\,>\,y_1$
in the first hadron ( see Fig.2a ) and all partons  with
$y_j\,<\,y_2$ in the second hadron  do not interact inelastically
and, hence,  do not  produce additional hadrons in the LRG. This
is a  difficult problem, since  not only partons at short
distances contribute to such a calculation, but also partons at long
distances for which  the pQCD approach is not valid.
Many attempts have been made to estimate $<
\mid S_{spectator}(\,s\,) \mid^2 >
$\cite{Bj}\cite{GLM1}\cite{ELSP}\cite{FLET}\cite{RR}\cite{GLMSM}\cite{GLMSP2}, 
but  this problem still
awaits a solution.
\end{itemize}

On the other hand, the experimental studies of LRG processes
have progressed,  and  have yielded  interesting results
 both at the Tevatron\cite{CDF}\cite{D0}, and at HERA\cite{DERR}.

The main results  of the experimental data pertaining to  the survival
probability are:
\begin{itemize}
\item\,\,\, the value of $<\mid S \mid^2 >$ is rather small. Indeed, the
experimental values  for
$f_{gap}(\, \sqrt{s}\,=\,630\,GeV\,)\,=\,1.6\,\pm 0.2\%$
 ( D0\cite{D0}) $=\,2.7\,\pm 0.9\%$ ( CDF\cite{CDF} )
and $ f_{gap}(\, \sqrt{s}\,=\,1800\,GeV
\,)\,=\,0.6\,\pm\,0.2\%$  ( D0\cite{D0} ) $=   1.13 \,\pm\, 0.16 \%$
 ( CDF\cite{CDF} ) can  be understood only if $<\mid S
\mid^2 >\,\,\approx\,\,1 - 10 \%$;

\item\,\,\, the energy dependence of $f_{gap}$, namely, the obsevation
that\cite{CDF}\cite{D0}
$$ R^{CDF}_{gap}\,\, =\,\, \frac{f_{gap}(\sqrt{s} = 630\; GeV)}
{f_{gap}(\sqrt{s}\;=\; 1800\; GeV)} \;\;= \;\; 2.4 \;\;\pm\,\, 0.9$$
$$  R^{D0}_{gap}\,\, = \,\,\frac{f_{gap}(\sqrt{s} = 630\; GeV)}
{f_{gap}( \sqrt{s}\;=\; 1800\; GeV)} \;\;= \;\; 2.67 \;\;\pm\,\,
0.38$$
leads to  $<\mid S \mid^2 >$ which decreases by a factor two in the above
range of energy. So, we expect that
$$ R_{S}\,\,=\,\,\frac{< \mid S \mid^2>_{\sqrt{s}=630}}
{< \mid S\mid^2>_{\sqrt{s}=1800}} \;\;\approx \;\; 2\,\,. $$
A recent D0 estimate\cite{D0}    of the above ratio is $ 2.2 \,\pm\,0.8$.
\end{itemize}

It was shown in Ref.\cite{GLMSP2} that the Eikonal Model for the
``soft" interaction at high energies,  is able to describe the
 features
of the experimental data. However, we may
question  the  reliability  of this approach. Especially
worrying,  is   the energy dependence of $ < \mid
S\mid^2>$,  since a natural parameter related to the parton
rescatterings is the ratio
\beq \label{I4}
R_D\,\,\,=\,\,\,\frac{\s_D}{\s_{tot}}\,\,\,\equiv\,\,\,\frac{\s_{el}\,\,+\,\,
\s_{SD}\,\,+\,\,\,\s_{DD}}{\s_{tot}}\,\,,
\eeq
where $\s_{SD}$ and $\s_{DD}$ are the cross sections of single and
double diffraction.  Experimentally,
$R_D$ is approximately constant over  a wide range of
energy. In the Eikonal Model we consider $\s_{SD}\,\,\ll\,\,\s_{el}$ and
$\s_{DD}\,\,\ll\,\,\s_{el}$ and, therefore, we model
$R_D\,\,\ra\,\,R_{el}\,\,=\,\,\frac{\s_{el}}{\s_{tot}}$.
Experimentally, $R_{el}$  depends    on energy, which
 gives rise  to  a considerable decrease of the survival probability.
Indeed, the first attempt to take into account the parameter $R_D$ in the
calculation of $ < \mid S\mid^2>$ given in Ref.\cite{RR}, leads to
\beq \label{I5}
< \mid S\mid^2>\,\,=\,\,[\,1\,\,-\,\,2\,\,R_D\,]^2\,\,,
\eeq
which yields a reasonable  value of the survival probability, but
cannot
account for
its substantial dependence on energy  ( see Ref.\cite{GLMSP2} ).

The main goal of this paper is to develop a  simple model for the ``soft"
high energy interactions, which correctly includes   the processes
of  diffractive dissociation, and to study the value and energy
dependence of $< \mid S_{spectator}( \,s\,) \mid^2>$. In the next section
we develop our approach, while in section 3 we discuss the value of  $<
\mid S_{spectator}( \,s\,) \mid^2>$ in a  model which describes the
energy dependence of $R_{el}$ and $R_D$. We show that our model  gives
the experimentally observed decrease of the survival probability as a
function of energy. In section 4 we discuss and summarize our results.

\section{A  three channel model}

\subsection{Diffractive dissociation ( general approach ) }
Diffractive dissociation is the simplest process with a  LRG in
which no hadrons are produced in  the central rapidity region.
In these processes we have
    production
of one
( single diffraction (SD) ) or
two groups of hadrons   ( double diffraction ( DD ) ) with masses (
$ M_1$
and $ M_2$ in Fig.3 ) much less than the total energy (
$  M_1\,\ll\,\sqrt{s} $ and
$ M_2\,\ll\,\sqrt{s} $ ).

\begin{figure}
\centerline{\psfig{file= 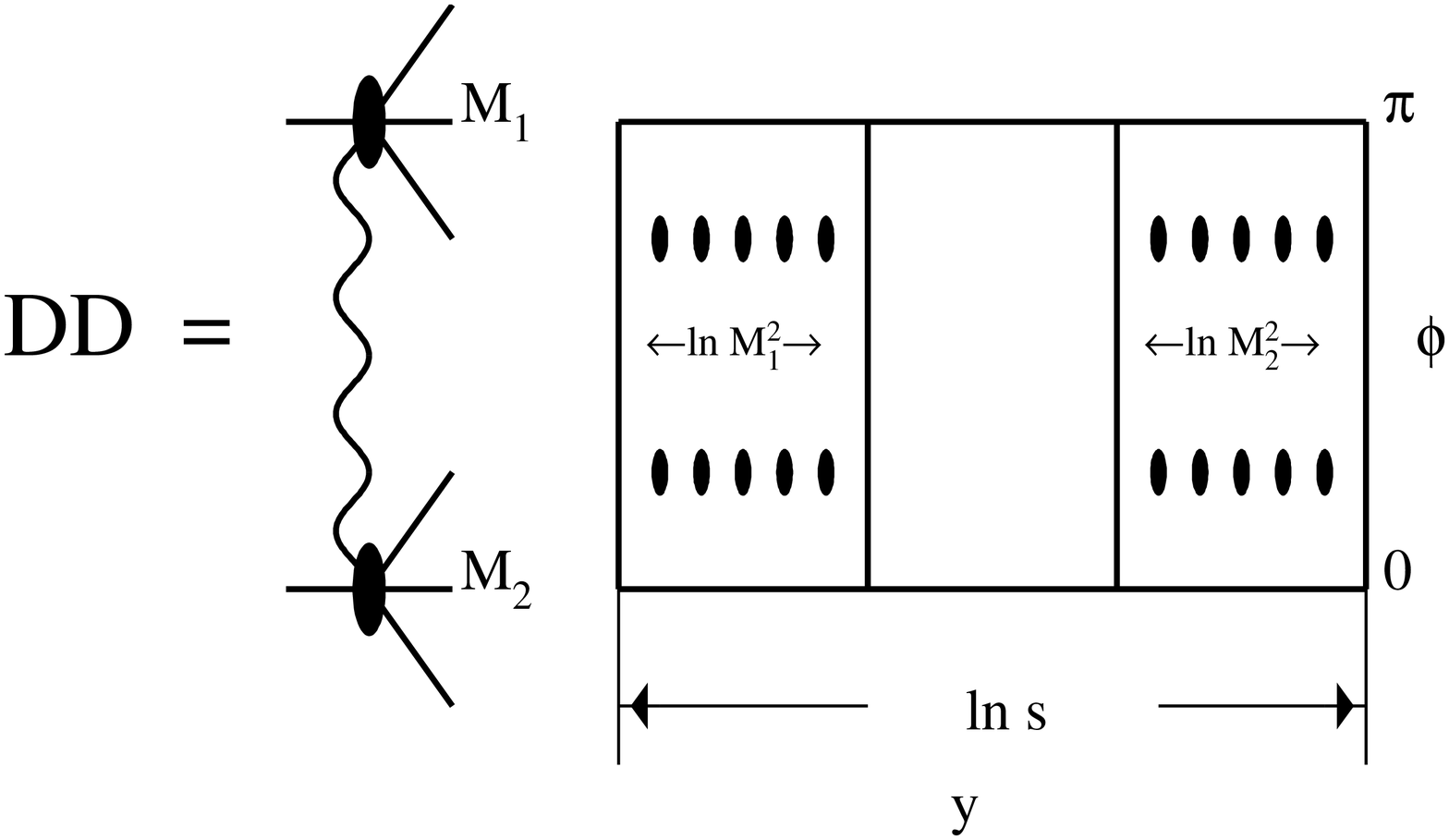,width=100mm,height=60mm}}
\caption{ \it Lego - plot for double diffractive dissociation.}
\label{fig3}
\end{figure}

 From  a theoretical point of view, as was suggested by
Feinberg\cite{FEIN} and Good and Walker\cite{GW}, diffractive
dissociation can be viewed as
{ \it a typical quantum mechanical process which occurs
since the hadron states are not diagonal with respect to the strong
interaction scattering matrix.}

 We    consider this point in more detail, and denote the
wave
functions which are diagonal with respect to the strong interaction by
$\Psi_{n}$. The
quantum numbers $n$ are called
the correct degrees of
freedom at high energy.
The amplitude  of the high energy interaction is, therefore,
given by
\beq \label{DD1}
A_{n_1 n_2}\,\,=\,\,< \Psi_{n_1} \Psi_{n_2} | T | \Psi_{n_1'} \Psi_{n_2'}
>\,\,=\,\,A_{n_1, n_2}\,\delta_{n_1,n_1'}\,\delta_{n_2,n_2'}\,\,,
\eeq
where the brackets denote all needed integrations and $T$ is the
scattering
matrix.  

The wave function of a hadron is
\beq \label{DD2}
 \Psi_{hadron}\,\,\,=\,\,\,\sum^{\infty}_{n =1} \,C_{n}\,
\Psi_{n}\,\,.
\eeq
For a hadron - hadron interaction we have a wave function
$\Psi_{hadron}\,\times\,
\Psi_{hadron}$ before the  collision, while after the  collision   the
scattering
matrix $T$ gives a new wave function
\bea \label{DD3}
&\Psi_{final}\,\,\,= &\nonumber\\
&\sum^{\infty}_{n_1 =
1}\,\sum^{\infty}_{n_2=1}\,\sum^{\infty}_{n_1'=1}\,\sum^{\infty}_{n_2'=1}\,\,
C_{n_1}\,C_{n_2}
< \Psi_{n_1}\,\Psi_{n_2} | T | \Psi_{n_1'}\,\Psi_{n_2'}>
\Psi_{n_1'}\,\Psi_{n_2'}\,\,= &\nonumber\\
&\,\,\sum^{\infty}_{n_1 =
1}\,\sum^{\infty}_{n_2 = 1}\,\, C_{n_1}\,C_{n_2}
\,A_{n_1,n_2}\,\Psi_{n_1}\,\Psi_{n_2}\,\,.&
\eea

 \eq{DD3} leads to the  elastic amplitude
\beq \label{DD4}
a_{el}\,\,=\,\,< \Psi_{final} | \Psi_{hadron}\,\times\,\Psi_{hadron}
>\,\,=\,\,\sum^{\infty}_{n_1 =1}\,\,\sum^{\infty}_{n_2
=1}\,\,C^2_{n_1}\,\,C^2_{n_2}\,\, \,A_{n_1,n_2}\,\,,
\eeq
and to another process, namely, to the production of other hadron states,
since
$\Psi_{final}$  may be different from
 $\Psi_{hadron}\,\times\,\Psi_{hadron}$.
\bea \label{DD5}
&\s_{D}(s,b)\,\,=\,\,< \Psi_{final} | \Psi_{final}>
\,\,-\,\,< \Psi_{final} | \Psi_{hadron}\,\times\,\Psi_{hadron} >^2\,\,=
&\nonumber\\
&
\sum^{\infty}_{n_1 = 1}\,\sum^{\infty}_{n_2 = 1}\,\,C^2_{n_1}\,C^2_{n_1}
\,A^2_{n_1,n_2}\,\,-\,\,\left(\,\sum^{\infty}_{n_1 =1}\,\,\sum^{\infty}_{n_2
=1}\,\,C^2_{n_1}\,\,C^2_{n_2}\,\, \,A_{n_1,n_2}\,\,\right)^2\,\,.&
\eea
 Using the normalization condition for the hadron wave function
( $ \sum_n \,C^2_n\,\,=\,\,1 $ ),  the cross
section of the diffractive dissociation processes, \eq{DD5},   can be
reduced
to the form\cite{GW}\cite{PUM}
\beq \label{DD6}
\s_{D}(s,b)\,\,=\,\,< | \s^2(s,b) | >\,\,-\,\,<| \s (s, b ) |
>^2\,\,,
\eeq
where $<| f | >\,\,\equiv\,\,\sum_{n_1}\,\,
\sum_{n_2}\,\,C^2_{n_1}\,C^2_{n_2}\,\, f_{n_1,n_2}$ and   we have
returned  to  our original  variables: energy ( $s$ ) and
impact parameter
( $b$ ).

\subsection{Diffractive dissociation and shadowing corrections}

For  $A_{n_1,n_2}(s, b)$,  and only for  $A_{n_1,n_2}(s, b)$, we have
the unitarity constraint
\beq \label{SC1}
2\,\,Im A^{el}_{n_1,n_2} (s, b )\,\,=\,\,| A^{el}_{n_1,n_2} (s, b
)|^2\,\,
+\,\,G^{in}_{n_1,n_2} (s, b)\,\,.
\eeq
Assuming that the amplitude at high energy is predominantly imaginary, we
obtain the solution of \eq{SC1}
\bea
&
A^{el}_{n_1,n_2}(s, b )\,\,\,=\,\,\,i\,\left( \,1\,\,\,-\,\,\,e^{ -
\,\frac{\O_{n_1,n_2}(s,b)}{2}}\,\right)\,\,;&\label{SC2}\\
&
G^{in}_{n_1,n_2} (s, b)\,\,\,=\,\,\,1\,\,\,-\,\,\,e^{ -
\,\O_{n_1,n_2}(s,b)}\,\,.&\label{SC3}
\eea
To find a relation between the processes of diffraction dissociation
and the value of the shadowing corrections ( SC ), we assume that
$\O_{n_1,n_2} ( s, b)\,\,\ll\,\,1$ and expand \eq{SC1} - \eq{SC3} with
respect to $\O_{n_1,n_2}$
\begin{eqnarray}
&
A^{el}_{n_1,n_2}(s,b)\,\,=\,\,\frac{\Omega_{n_1,n_2}(s,b)}{2}\,\,\,-
\,\,\frac{\Omega^2_{n_1,n_2}(s,b)}{8}\,\,\,+\,\,\,O
(\Omega^3_{n_1,n_2})\,\,;
& \label{SC4}\\
&
G^{in}_{n_1,n_2}(s,b)\,\,\,=\,\,\,\Omega_{n_1,n_2}(s,b)\,\,\,-\,\,\,
\frac{\Omega^2_{n_1,n_2}(s,b)}{2}\,\,\,+\,\,\,O
(\Omega^3_{n_1,n_2})\,\,.& \label{SC5}
\end{eqnarray}
Using \eq{DD2} - \eq{DD5} we obtain for the  observables
\begin{eqnarray}
&
\s_{el}(s,b)\,\,\,=\,\,\,\frac{1}{4}\,\,
\left(\,\sum^{\infty}_{n_1=1}\,\sum^{\infty}_{n_2=1}\,C^2_{n_1}\,C^2_{n_2}
\,\Omega_{n_1,n_2}(s,b)\,\right)^2
& \label{SC6}\\
&
\s_{tot}\,=\,\sum^{\infty}_{n_1=1}\,\sum^{\infty}_{n_2=1}\,\,
C^2_{n_1}\,C^2_{n_2}\,\Omega_{n_1,n_2}(s,b)
\,-\,\frac{1}{4}\,
\left(\,\sum^{\infty}_{n_1=1}\,\sum^{\infty}_{n_2=1}\,\,
\,C^2_{n_1}\,C^2_{n_2}\,\Omega^2_{n_1,n_2}(s,b)\,\right)&
\label{SC7}\\
&
\s_{in}\,=\,\sum^{\infty}_{n_1=1}\,\sum^{\infty}_{n_2=1}\,\,
\,C^2_{n_1}\,C^2_{n_2}
\Omega_{n_1,n_2}(s,b)
\,- \,\frac{1}{2}\,\left(\,\sum^{\infty}_{n_1=1}\,\sum^{\infty}_{n_2=1}
\,\,
\,C^2_{n_1}\,C^2_{n_2}
\Omega^2_{n_1,n_2} (s,b)\,\right)& \label{SC71}\\
&
\s_{diff}\,=\,\frac{1}{4}\,\{\sum^{\infty}_{n_1=1}\,
\sum^{\infty}_{n_2=1}
\,C^2_{n_1}\,C^2_{n_2}
\,\Omega^2_{n_1,n_2}(s,b) - \left(\,
\sum^{\infty}_{n_1=1}\,\sum^{\infty}_{n_2=1}
\,C^2_{n_1}\,C^2_{n_2}
\Omega_{n_1,n_2}(s,b)\,\right)^2\}. &\label{SC8}
\end{eqnarray}

In  the
parton model,  one Pomeron exchange corresponds to a typical
inelastic event with a production of a large number of particles.
In this case, we can associate this exchange with $\Omega$,
consequently $\s^P_{tot}(s,b
)\,\,=\,\,\s^P_{in}( s, b )\,\,\propto \,\,\Omega$.
All terms which are proportional to $\Omega^2$ describe  two Pomeron
exchange, and  they induce  SC.

We can evaluate the scale of the SC using experimental data on $
\s_{tot},\s_{el}$ and $ \s_{SD}$.
Indeed, we can write the expression for the total cross section in the
form
\beq \label{SC9}
\s_{tot}\,\,\,=\,\,\s^P_{tot}\,\,\,-\,\,\,\Delta \s^{SC}_{tot}\,\,,
\eeq
where $\s^P_{tot} $ is the contribution of  Pomeron exchange to the
total cross section.  Summing \eq{SC6} and \eq{SC8} we derive  that
\beq \label{SC10}
\Delta
\s^{SC}_{tot}\,\,=\,\,\s_{el}\,\,+\,\,\s_{diff}\,\,=\,\,\s_{el}\,\,+\,\,
\s_{SD}\,\,+\,\,\s_{DD}\,\,=\,\,\s_{D}\,\,.
\eeq
The ratio $R_D$ ( see \eq{I5} ) gives the scale of the
SC
in the situation when these are sufficiently weak  ( $\O_{n_1,n_2}
\,\,\ll\,\,1 $
), since \eq{SC10} can be rewritten in the form
\beq \label{SC11}
R_D\,\,=\,\,\frac{\D\,\s^{SC}_{tot}}{\s_{tot}}\,\,.
\eeq
 Fig.4b  shows that over a wide range of energy $R_D
\,\,\approx\,\,0.34$ , and it appears to be independent of energy.
The
large value of $R_D$ implies that  SC should be taken into
account,  and that  SC lead to  the small value of the survival
probability. The almost constant $R_D$ suggests that
the SC cannot induce the observed strong energy dependence of the survival
probability. However, the value of $R_D$ is so large that we have to
develop an approach to calculate  SC for
$\O_{n_1,n_2}\,\,\approx\,\,1$.

 \begin{figure}
\begin{tabular} {c c}
\epsfig{file=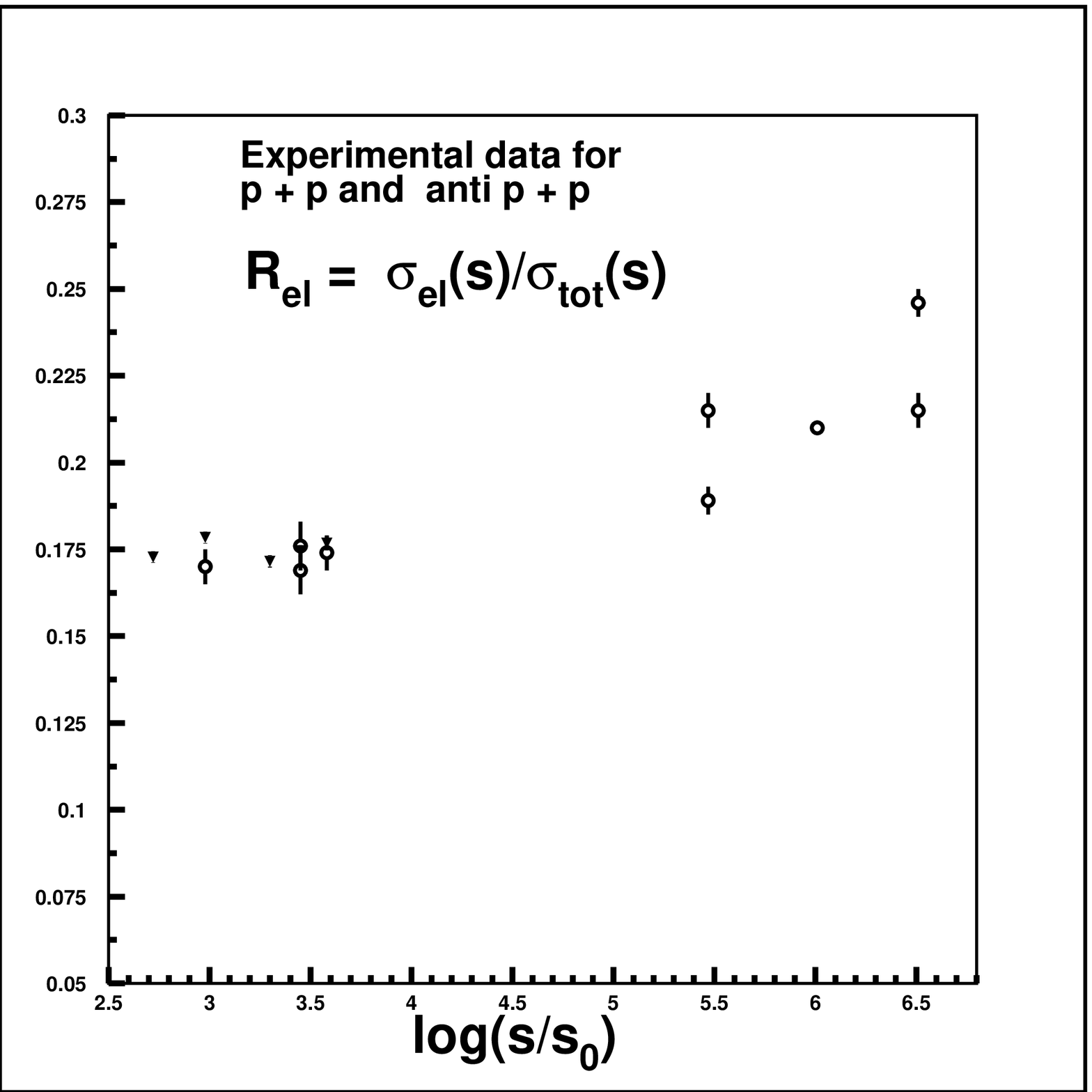,width= 80mm} &
\epsfig{file=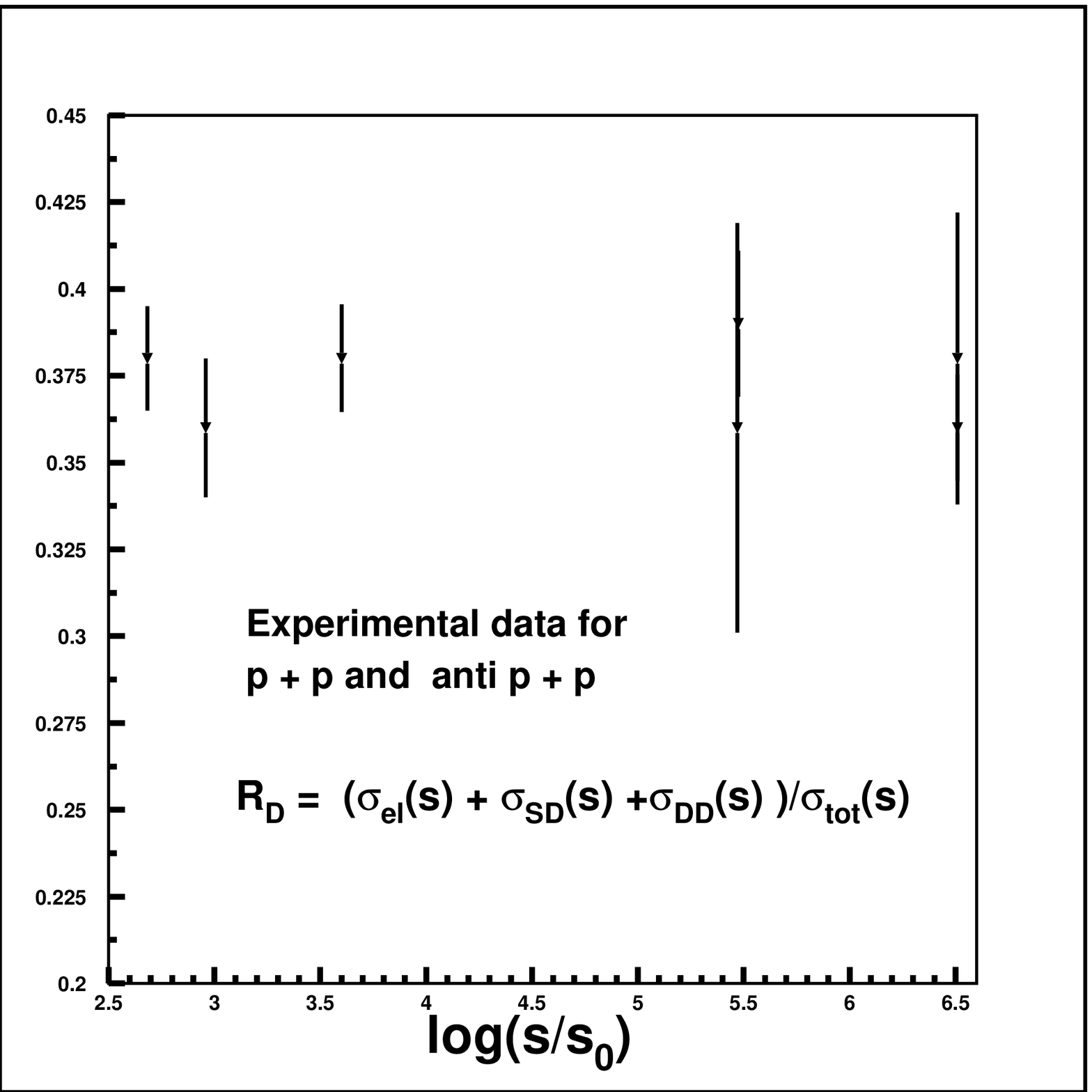,width=80mm}\\
Fig. 4-a & Fig.4-b\\
\end{tabular}
\caption{\it The experimental data on the ratio $R_{el}
\,=\,\s_{el}/\s_{tot}$ ( Fig.4a ) and on the ratio  $R_D\,=\,( \s_{el} +
\s_{SD} + \s_{DD} )/\s_{tot}$ ( Fig.4b ) versus $\log(s/s_0)$ with $s_0 =
1\,GeV^2$.}
\label{Fig.4}
\end{figure}

\subsection{The Eikonal Model}
We first discuss the Eikonal model. This is an approximation which
has been widely used to estimate the value of the SC, in a
situation when they are not small. The main assumption of this
model is that hadrons are the correct degrees of freedom at high
energy. In other words, we assume that the interaction matrix is
diagonal with respect to hadrons. From \eq{DD2} - \eq{DD6} one can
see that this model does not include diffractive dissociation
processes. Therefore, the accuracy of our estimates in the Eikonal
Model will be given by the ratio $ (\s_{SD} +
\s_{DD})/\s_{el}$.  From Fig.4 one can see that this ratio is about 1 at
$\sqrt{s}\,\,\approx\,\,20 \,GeV$ and decreases  reaching the value
of about 0.4 - 0.5 at the Tevatron energies. Thus,  we cannot
expect   the Eikonal approach to  yield  a reasonable estimate.
However, this model has the advantage of being simple, and  it
provides a good illustration, given below,  of the main
 elements and approximations used in previous calculations.

\begin{enumerate}

{\bf \item\,\,\, Assumptions:}
\begin{itemize}
\item\,\,\, Hadrons are the correct degrees of freedom at high energies\,;

\item\,\,\, $\s_{SD}\,\,\ll\,\,\s_{el}$ and $\s_{DD}\,\,\ll\,\,\s_{el}$\,;

\item\,\,\,At high energy the scattering amplitude is almost pure
imaginary $ Re \,a_{el}\,\,\ll\,\,Im \,a_{el}$\,;

\item\,\,\, Only the fastest partons can interact with each other\,\,.
\end{itemize}

The last assumption is the most restrictive, and  clearly indicates how
far
from
reality the Eikonal Model estimates could be.

{\bf \item\,\,\, Unitarity:}

In the Eikonal Model we only  have  one amplitude, since the scattering
matrix is diagonal in the hadronic wave functions. Therefore, the
unitarity
constraints of \eq{SC1} simplify to
\beq \label{EM1}
2\,\,Im \,a_{el} (s, b )\,\,=\,\,| a_{el} (s, b
)|^2\,\,
+\,\,G^{in} (s, b)\,\,.
\eeq
\eq{EM1} has the solution
\bea
&
a_{el}(s, b )\,\,\,=\,\,\,i\,\left( \,1\,\,\,-\,\,\,e^{ -
\,\frac{\O(s,b)}{2}}\,\right)\,\,;&\label{EM2}\\
&
G^{in} (s, b)\,\,\,=\,\,\,1\,\,\,-\,\,\,e^{ -
\,\O(s,b)}\,\,.&\label{EM3}
\eea

{\bf \item\,\,\, The Pomeron hypothesis:}

The main assumption of the Eikonal Model  is the identification  of
the opacity $\O(s,b)$ with a  single Pomeron exchange, namely
\begin{eqnarray}
&
\Omega(s,b)\,=\,\Omega^P(s,b)\,\,; &\label{EM4}\\
&
\Omega^P(s,b)\,\,=\,\, \s_0\,\G(b)\,\,=\,\,\frac{\sigma_0}{\pi R^2(s)}\,(
\frac{s}{s_0}
)^{\Delta_P}\,e^{-\frac{b^2}{R^2(s)}}\,\,
=\,\,\nu(s)\,\,e^{-\frac{b^2}{R^2(s)}}\,;&\label{EM5}\\
 &
R^2(s)\,=\,\,4\,R^2_0\,\,+\,\,4\alpha'_P\,\ln(s/s_0)\,\,; &\label{EM6}\\
&
\nu(s)\,\,=\,\,\frac{\sigma_0}{\pi R^2(s)}\,( \frac{s}{s_0}
)^{\Delta_P}\,\,=\,\,\Omega^P(s,b = 0 )\,\,. & \label{EM7}
\end{eqnarray}
We assume a Pomeron trajectory
$\alpha_P(t)\,=\,\alpha_P(0)\,\,+\,\,\alpha_P'\,t$.
 \eq{EM4} is a reasonable approximation  in the kinematic region  where
$\Omega$ is
small, i.e.  either at low  energies or at high
energies when  $b$ is large. Therefore, the Eikonal
approach is the natural generalization  of the single Pomeron exchange
satisfying   $s$-channel
unitarity. \eq{EM4} is an  explicit
analytic expression for the  well known partonic picture for the
Pomeron structure, namely, the single Pomeron exchange is
responsible for the inelastic production of particles which are
uniformly distributed in  rapidity.

{\bf \item\,\,\, Exponential parameterization: }

In  \eq{EM5} a Gaussian form is  explicitly  assumed for the profile
function
$\G(b)$. This corresponds to an exponential form in $t$ space.

\beq \label{EM8}
\G(b)\,\,\,=\,\,\frac{1}{\pi\,R^2(s)}\,\,e^{-\frac{b^2}{R^2(s)}}\,\,.
\eeq

This form is  assumed  due to its simplicity, the resulting integrals
can be done analytically and  we can write the explicit  answer
for the
physical observables. The proper  definition of the profile function
$\G(b ) $ is given by the following  Fourier transform
\beq \label{EM9}
\G(b)\,\,\,=\,\,\frac{1}{2 \pi} \int\,\,d^2 b\,\,e^{ - i\mathbf{
\vec{q}\,\cdot \vec{b}}}\,\,G^2_P( q^2)\,\,,
\eeq
where $G_P(q^2 )$ is the Pomeron  form factor. For example, one can take
for $G_P (q^2 )$ the prediction of the Additive Quark Model ( see
Ref.\cite{DL} ) in which $G_P(q^2)\,\, = \,\, G_{em}$, where $G_{em}$ is
the electromagnetic form factor of a hadron.

{\bf \item\,\,\, $\mathbf{R_D}$ :}

Using \eq{EM4} - \eq{EM7} one can obtain a  closed expression for the
``soft" observables

\begin{eqnarray}
&
\sigma_{tot}\,\,=\,\,2\,\pi
\,R^2(s)\,\{\,\ln(\nu(s)/2)\,\,+\,\,C\,\,-\,Ei( - \nu/2)
\,\}\,\,; & \label{EM10}\\
&
\sigma_{in}\,\,=\,\,\,\pi
\,R^2(s)\,\{\,\ln(\nu(s))\,\,+\,\,C\,\,-\,Ei( - \nu)\,\}\,\,; &
\label{EM11}\\
&
\sigma_{el}(s)\,\,=\,\,\sigma_{tot} (s)
\,\,-\,\,\sigma_{in}(s)\,\,; & \label{EA7}
\end{eqnarray}
where
$ Ei(x)\,\,\,=\,\,\int^{x}_{- \infty}\,\,\frac{e^{t}}{t} \,dt $
and $C \,\,=\,\,0.5773$.

We define
\beq \label{EM12}
R_{el}\,\,=\,\,\frac{\s_{el}}{\s_{tot}}\,\,=\,\,\frac{\ln[\frac{\nu}{4}]\,\,+
\,\,C\,\,+\,\,Ei(
- \nu )\,\,-\,\,2\,Ei( -
\frac{\nu}{2})}{2\,[\,\ln[\frac{\nu}{2}]\,\,+\,\,C\,\,- \,\,Ei( -
\frac{\nu}{2})\,]}\,\,.
\eeq
Note, that the ratio $R_{el}$ depends only on $\nu$ and does not depend on
the value of radius.
Using \eq{EM12} one can find the value of $\nu$ from the
experimental data on $R_{el}$ ( see Fig.4a ) this was  done in
Ref.\cite{GLMSP2}. With the  value of $\nu$  determined in this way we
can 
calculate the survival probability.

{\bf \item\,\,\, Survival probability :}

From \eq{EM3}  one can conclude that the  factor
\beq \label{EM13}
P(s,b)\,\,\,=\,\,\,e^{ - \O(s,b)}\,\,,
\eeq
  is  the probability that the two initial hadrons
do not interact inelastically. In a  QCD
approach, this  means that the fastest  parton from one hadron does not
interact with the fastest parton from another. Therefore, in the
Eikonal Model the survival probability can be easily calculated in
the following way\cite{Bj}\cite{GLM1}
\beq \label{EM14}
< \mid S_{spectator}(\,s\,) \mid^2>\,\,\,=\,\,\,\frac{\int\,\,d^2 b
\,P(s,b)\,A_{HP}(\Delta y,b)}{\int\,\,d^2 b A_{HP}(\Delta y,b)}
\,\,,
\eeq
where $A_{HP}( \Delta y, b )$ is the  cross section  for a two parton
jet
 production with a  LRG due to single ``hard" Pomeron exchange ( see Fig.
1 ). It has been proven that for a  ``hard" cross section, the $b$
dependence can be factorized out\cite{ELSP}\cite{GLR}\cite{GLMF2SL}. 
If we assume
$A_{HP}(\Delta y, b )$ to be Gaussian we have

\beq \label{EM15}
A_{HP}(\Delta y, b )\,\,\,=\,\,\,\sigma_{HP} ( \Delta y )\,\G_H (b)
\,\,\,=\,\,\frac{\sigma_{HP} ( \Delta y )}{\pi R^2_H}\,\,e^{ -
\frac{b^2}{R^2_H}}\,\,,
\eeq
 where $R^2_H$ is  the radius of the ``hard" interaction.

Based on  this assumption we  obtain for the survival
probability:
\beq \label{EM16}
< \mid S_{spectator}(\,s\,) \mid^2> =  \frac{a \gamma[a,
\nu]}{\nu^{a}}\,\,,
\eeq
where the incomplete gamma function
$ \gamma(a,x) = \int_{0}^{x} z^{a-1}e^{-z}dz $
and
$a\,\,=\,\,\frac{R^2(s)}{R^2_H}\,\,.$

\end{enumerate}

In Ref.\cite{GLMSP2}, \eq{EM12} and \eq{EM16} were used to calculate
the value and energy dependence of the survival probability. It was shown
that both the value and the energy dependence are sensitive to the value
of the ``hard" radius, which was extracted in Ref.\cite{GLMSP2} from the
experimental data on (i) vector meson diffractive dissociation in DIS at
HERA\cite{HERADDDATA},
 and
on (ii) the CDF double parton cross section at the Tevatron\cite{CDFDP}.

Even though the Eikonal Model, as used in Ref.\cite{GLMSP2},
can reproduce both the experimentally measured   value and its energy
behaviour,
 the  reliability of such an  approach is questionable. In the following
we
attempt to construct a more realistic model.

\subsection{ A three channel model: assumptions and general formulae}

We want to construct  a model which takes into account the processes of
diffractive dissociation, for  the case when these are not small (
$\s_{SD}\,\,\approx\,\,\s_{el} $ ).

\subsubsection{Assumptions:}

The main idea behind  the model which is  presented  here, is to
replace the many  final states of the diffractively produced hadrons by
one
state ( effective hadron ). Doing so, we assume that we have two wave
functions which are diagonal
with respect to the strong interactions: $\Psi_1$ and $\Psi_2$. In this
case the general
\eq{DD2}  can be reduced to the form
\beq \label{QE1}
\Psi_{hadron}\,\,\,=\,\,\,\alpha\,\Psi_1\,\,\,+\,\,\,\beta
\,\Psi_2\,\,\,,
\eeq
with the condition $\a^2\,+\,\b^2\,\,=\,\,1$, which follows from
the
normalization of the wave function.
The wave function of the diffractively produced bunch of hadrons should be
orthogonal to $\Psi_{hadron}$ and has the form
\beq \label{QE2}
\Psi_D\,\,\,=\,\,\,-\,\beta\,\Psi_1\,\,\,+\,\,\,\a\,\Psi_2\,\,.
\eeq
\eq{QE2} is the explicit form of our assumption,  
 that  we replace the complicated
final state of  diffractively  produced systems of hadrons 
by one wave function $\Psi_D$.

\subsubsection{General formulae:}

Substituting  \eq{QE1} and \eq{QE2} in \eq{DD3}  we can obtain
 \beq \label{QE3}
\Psi_{final}\,\,\,=\,\,\a^2\,A_{1,1}\Psi_1\,\times\,\Psi_1
\,\,+\,\,\a\,\b
\,A_{1,2}\,[\,\Psi_1\,\times\,\Psi_2\,\,+\,\,\Psi_2\,\times\,
\Psi_1\,]\,\,+\,\,\b^2\,A_{2,2}\,
\Psi_2\,\times\,\Psi_2\,\,.
\eeq
The amplitudes $A_{i,k}\,\,( i,k = 1,2 ) $     can be written in the form
of     \eq{SC2} and \eq{SC3} ( see Fig.5 ) 
\bea
&
A^{el}_{i,k}(s,b)\,\,=\,\,i\,\{\,\,1\,\,\,-\,\,\,e^{ -
\frac{\O_{i,k}( s, b )}{2}}\,\,\}\,\,; &\label{QE4}\\
&
G^{in}_{i,k}( s, b )\,\,\,=\,\,\,1\,\,\, - \,\,\,e^{ - \O_{i,k}( s, b
) }\,\,.&\label{QE5}
\eea
The elastic amplitude is equal to
$$
a_{el} (s, b)\,\,=\,\,<\, \Psi_{hadron}\,\times\, \Psi_{hadron}\, | T |\,
\Psi_{hadron}\,\times\, \Psi_{hadron} \,>\,\,=\,\,<\,
\Psi_{hadron}\,\times\,
\Psi_{hadron} | \Psi_{final} \,>\,\,=
$$
\beq \label{QE6}
\a^4\,A_{1,1} \,\,+\,\,2\,\a^2\,\b^2
\,A_{1,2}\,\,+\,\,\b^4\,\,A_{2,2}\,\,.
\eeq
For  single diffraction we have the amplitude
$$
a_{SD} (s, b)\,\,=\,\,<\, \Psi_{hadron}\,\times\, \Psi_{hadron}\, | T |\,
\Psi_D \, \times \,\Psi_{hadron} \,>\,\,=\,\,<\, \Psi_{hadron}\,\times\,
\Psi_D | \Psi_{final} \,>\,\,=
$$
\beq \label{QE7}
\a\,\b\,\{\,\,-\,\a^2\,A_{1,1} \,\,+\,\,(\,\a^2\,-\,\b^2\,)
\,A_{1,2}\,\,+\,\,\b^2\,\,A_{2,2}\,\,\}\,\,,
\eeq
while the amplitude for double diffractive production is

\beq \label{QE8}
a_{DD} (s, b)\,\,=\,\,<\, \Psi_{hadron}\,\times\, \Psi_{hadron}\, | T |\,
\Psi_D \,\times\, \Psi_D \,>\,\,=\,\,<\, \Psi_D\,\times\,
\Psi_D | \Psi_{final} \,>\,\,=
\eeq
$$
\a^2\,\b^2\,\{\,\,A_{1,1} \,\,-\,\,2
\,A_{1,2}\,\,+\,\,A_{2,2}\,\,\}\,\,.
$$
\eq{QE6} - \eq{QE8} together with \eq{QE4} and \eq{QE5} give the general
formulae for our model.

\begin{figure}
\centerline{\epsfig{file=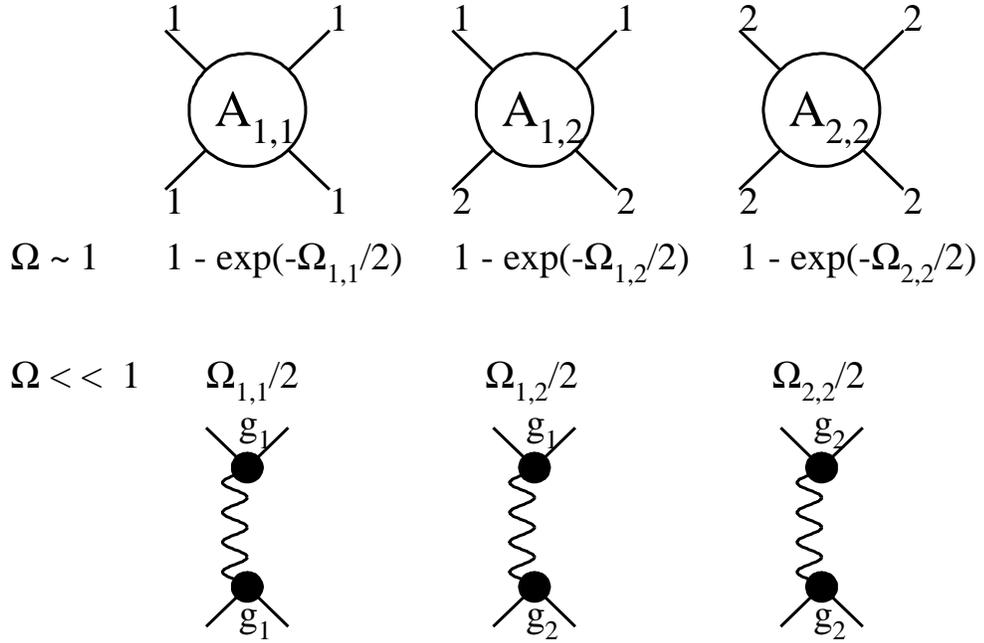,width=150mm}}
\caption{ \it Solution of the unitarity constraints and the Regge
parametrization for the correct degrees of freedom in the  three channel
model. The waved lines denote the exchange of the Pomeron.}
 \label{Fig.5}
\end{figure}

In general this,  three channel model is an attempt to sum all
rescatterings shown in Fig.6. All of them are eikonal type
  rescatterings, but  contrary to the Eikonal Model, the quasi-elastic
rescatterings with production and rescattering  of an effective
diffractive state   has been taken into account.

\begin{figure}
\centerline{\epsfig{file=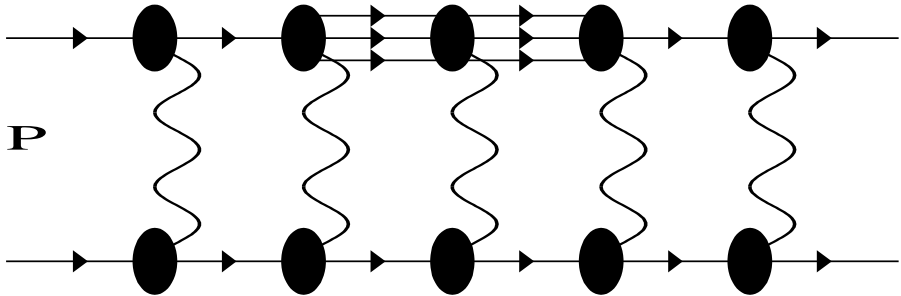,width=160mm,height=55mm}}
\caption{ \it The Pomeron interaction in the three channel model.}
 \label{Fig.6}
\end{figure}

We call this model a  three channel one because, three physical
processes:
elastic  scattering, single and double diffraction, are included in it.
All general formulae of \eq{DD1} - \eq{DD6} were known  long ago ( see
Ref.\cite{VL}, for example ). In  Ref.\cite{FLET}  this general
 formalism was applied to obtain  estimates of the value of
the survival probability. We  now  develop a systematic
analysis to obtain  the value of $< \mid S^2 \mid >$ in the framework of
the three
channel  model, utilizing the experimental data pertaining to the
``soft" processes that have been measured.

\subsection{ A three channel model: physical observables}
For the opacities $\O_{i,k}$ we use the same approach as in the Eikonal
Model
( see \eq{EM4} - \eq{EM7} and Fig. 5 )
 \beq \label{PO1}
\O_{i,k}\,\,=\,\,\nu_{i,k}(s) \,e^{ -\frac{b^2}{R^2_{i,k}(s)}}
\,\,=\,\,\frac{g_k g_i}{\pi
\,R^2_{i,k}(s)}\,\left( \,\frac{s}{s_0}\,\right)^{\Delta_P}\, e^{
-\frac{b^2}{R^2_{i,k}(s)}}\,\,,
\eeq
where we have used  Reggeon factorization as it is shown in Fig. 5.
 In this paper,  we do not  use the exact Reggeon dependence on energy,
 but we  utilize the factorization properties which appear  in
\eq{PO1}. In addition we need to take into account the
factorization properties of the radii
\beq \label{PO2}
R^2_{i,k}\,\,=\,\,2\,R^2_{i0}\,\,+\,\,2\,R^2_{k0}\,\,+
\,\,4\,\alpha'_P\,\ln(s/s_0)\,\,,
\eeq
where $R^2_{i0}$ is a radius which describes the $t$-dependence of the
Pomeron - hadron $i$  vertex. It is obvious from \eq{PO1} and
\eq{PO2}
that $\O_{i,k}$ can be written in the form
\bea
&
\O_{1,1}\,\,=\,\,\nu_1\,X\,\,; &\label{PO3}\\
&
\O_{2.2}\,\,=\,\,\nu_2\,X^{\frac{r}{2 - r}}\,\,;&\label{PO4}\\
&
\O_{1,2}\,\,=\,\,\sqrt{\nu_1\,\nu_2\,r\,(2 - r )}\,X^r;&\label{PO5}
\eea
where $X\,\,=\,\,e^{ -\frac{b^2}{R^2_{1,1}(s)}}$ and
$r\,\,=\,\,\frac{R^2_{1,1}(s)}{R^2_{1,2}(s)}$.

\eq{PO3} - \eq{PO5} together with \eq{QE5} - \eq{QE8} allow us to express all
physical observables through the  variables $\nu_1$, $ \nu_2$,$r$ and
$\b$. The first three variables  depend on energy squared ( $s$ ) (
see \eq{PO1} and \eq{PO2} ) while $\b$ is a constant in our model.
From \eq{PO1}  we expect $\nu_2$ to be proportional
to $\nu_1$, and $r$ to be only weakly ( logarithmically ) dependent on
energy.

From \eq{PO3} - \eq{PO5} we have the following  expressions for the
amplitudes
$A_{i,k}$ in terms of  our variables
\bea
& A_{1,1} ( \nu_1,X )\,\,=\,\,i\,\left(\,1\,\,-\,\,e^{-
\frac{\nu_1\,X}{2}}\,\right)\,\,;&\label{PO6}\\
& A_{2,2}(\nu_2,r,X )\,\,=\,\,i\,\left(\,1\,\,-\,\,e^{-
\frac{\nu_2\,X^{\frac{r}{2 - r}}}{2}}\,\right)\,\,;&\label{PO7}\\
&A_{1,2}(\nu_1,\nu_2,r,X)\,\,=\,\,i\,\left(\,1\,\,-\,\,e^{- \,\frac{
\sqrt{\nu_1\,\nu_2\,r\,( 2 - r )}\,X^r}{2}} \,\right)\,\,.&\label{PO8}
\eea

After some simple calculations we  have
\bea
&
\s_{tot}\,\,=\,\,2\pi R^2_{1,1}\,\int^1_0\,\,\frac{d X}{X}\,\,\{\,
(1 - \b^2)^2\,A_{1,1}\,\,+\,\,2\,( 1 - \b^2 )
\,\b^2\,A_{1,2}\,\,+\,\,\b^4\,A_{2,2}\,\}\,\,;&\label{PO9}\\
&
\s_{el}\,\,=\,\,\pi\,R^2_{1,1}\,\int^1_0 \,\,\frac{d X}{X}\,\,\{\,
( 1 - \b^2 )^2\,A_{1,1}\,\,+\,\,2\,( 1 - \b^2 )
\,\b^2\,A_{1,2}\,\,+\,\,\b^4\,A_{2,2}\,\}^2\,\,;&\label{PO10}\\
&
\s_{SD}\,=\,\pi\,R^2_{1,1}\,(1 - \b^2) \,\b^2\,\,
\int^1_0\,\,\frac{d X}{X}\,\,\{ - ( 1 - \b^2 )\,A_{1,1}\,\,+\,\,
( 1 - 2 \,\b^2)\,A_{1,2}\,\,+\,\,\b^2\,A_{2,2}\,\}^2;
&\label{PO11}\\ &
\s_{DD}\,\,=\,\,\pi\,R^2_{1,1}\,( 1 - \b^2)^2\,\b^4\,\,\int^1_0\,\,
\frac{d X}{X}\,\,\{\,A_{1,1}\,\,-\,\,2\,A_{1,2}\,\,+\,\,A_{2,2}\,\}^2\,\,.
& \label{PO12}
\eea
One can see that the ratio $R_D$ as well as $R_{el}$, does not
depend on $R^2_{1,1}$. We will use these ratios to fix our
variables from the experimental data.

\subsection{A three channel model : survival probability}
To calculate the survival probability of the LRG in our three channel
model we  recall that the physical meaning of the factor
\beq \label{S1}
P_{i,k}\,\,=\,\,e^{- \O_{i,k}( s,b)}\,\,\equiv\,\,\{\,1
\,\,-\,\,A_{i,k}\,\}^2\,\,
\eeq
is, the probability that two hadronic states with quantum numbers
$i$ and $k$  scatter, without any inelastic interaction at given
energy $s$ and impact parameter $b$. Therefore, we have to
multiply the ``hard" cross section of their interaction by
$P_{i,k}$, and sum over all possible $i$ and $k$ for hadron - hadron
collisions to obtain the survival probability. The cross section for
 two jet  production with large transverse momenta and LRG, can
be reduced to the form
\beq \label{S2}
\s_{H}(s,b)\,\,=
\eeq
$$
(1 - \b^2 )^2 \nu_1\,\frac{R^2_{1,1}}{\tilde{R}^2_{1,1}}
\,e^{- \frac{b^2}{\tilde{R}^2_{1,1}}}\,\,+\,\,2\,( 1 - \b^2)\,\b^2
\,\sqrt{\nu_1\,\nu_2\,a\,( 2 - a )}\,\frac{ R^2_{1,2}}{\tilde{R}^2_{1,2}}
\,e^{ - \frac{b^2}{\tilde{R}^2_{1,2}}}\,\,+\,\,\b^4\,\nu_2\,
\frac{R^2_{2,2}}{\tilde{R}^2_{2,2}}\,e^{ -
\frac{b^2}{\tilde{R}^2_{2,2}}}
\,\,,
$$
where $\tilde{R}^2_{i,k}$  denotes the ``hard" interaction radius
of 
two states $i$ and $k$. Strictly speaking,
$\tilde{R}^2_{i,k}\,\,= R^2_{i,k} (s=s_0)$ but, really, we do not
know the value of $s_0$. However, we will show in the next section
that we are able to find the value of $\tilde{R}^2_{i,k}$ directly
from experimental data.

Finally, the survival probability of the LRG is 
\beq \label{S3}
< \mid S^2_{spectator}(\,s\,) \mid
>\,\,\,= \,\,\,\frac{N(s)}{D(s )}\,\,,
\eeq
where
\beq \label{S4}
N(s)\,\,=\,\,
\int^1_0\,\frac{d X}{X}
\eeq
$$
\{\,(1 - \b^2 )^2 \,
P_{1,1} \,\nu_1 \,a_{1,1}X^{a_{1,1}}\,\,+\,\,2 (1 - \b^2 )
\,\b^2 \,P_{1,2}\,\nu_{1,2}\,a_{1,2}\,
X^{a_{1,2}\,r}\,\,+\,\,\b^4\,
P_{2,2}\,\nu_2\,a_{2,2}\,X^{a_{2,2}\,r/(2-r)}
\,\}
$$
and
\beq \label{S5}
D(s)\,\,=\,\,
 \int^1_0\,\frac{d X}{X}
\eeq
$$
\{\,( 1 - \b^2
)^2\,\nu_1\,a_{1,1}\,X^{a_{1,1}}\,\,+\,\,2\,(1 - \b^2 )
\,\nu_{1,2}\,a_{1,2}\,X^{a_{1,2}\,r}\,\,+\,\,\b^4\,
\nu_2\,a_{2,2}\,X^{ a_{2,2}\,r/(2 - r)}\,\}\,\,,
$$
where we denote $\nu_{1,2} = \sqrt{\nu_1\,\nu_2\,r\,( 2 - r)}$ and
$a_{i,k}\,\,=\,\,\frac{R^2_{i,k}}{\tilde{R}^2_{i,k}}$. In the next
section we  will determine all parameters from the experimental data, and
will find the  value and the energy dependence which  are
typical for the survival probability in our model.

\section{Numerical evaluation of $\mathbf{< \mid S^2_{spectator} \mid>}$
 from the experimental data}

\subsection{ Fixing the parameters of the model}

Following the ideas of Ref.\cite{GLMSP2} we determine  all the
parameters of the model directly from the experimental data.

\begin{enumerate}

\item\,\,\,The most striking experimental fact is that $R_D$ is almost
independent on
 energy. $ R_D$  is rather big ( $ R_D \,\approx\,0.4 $ )
 ( see Fig. 4b ). We found that this energy behaviour, as well as the
value
 of $R_D$, allowed us to find the coefficient $t $ in the equation
 $\nu_2 \,=\,t\,\nu_1$.

\item\,\,\, The energy behaviour and the value of $R_{el}$ ( see Fig. 4a )
is used to determine  the value of $\nu_2$ at different energies, as has
been done in Ref.\cite{GLMSP2}.

\item\,\,\, Unfortunately, no reliable measurement has been performed
on the double diffractive cross section. We used the estimates of
Ref.\cite{DDDINO} for  the value of this cross section,  to
check that our choice of $\nu_1 $ and $\nu_2$ is not in a
contradiction with the value of the double diffraction cross section.

\item\,\,\, We used the experimental data on hard diffraction in DIS
to fix the value of the ``hard" radii in \eq{S3}.

\item\,\,\,Our goal is not to fix
all parameters, but rather to find out how  sensitive the value
and the  energy behaviour of the survival probability is,  to
uncertainties
in the values of the model parameters. We also   evaluate the range of
typical  values of  $< \mid S^2_{spectator}>$.
\end{enumerate}

\subsection{ $\mathbf{R_D}$ and $\mathbf R_{el} $ }

We start by  fixing  the model parameters  for  the   case of very high
energies for which $r\,\,\ra\,\,1$. In Fig.7 we plot $R_D $ and
$R_{el}$ at fixed $\b = 0.65$ versus $\nu_1$ and $t$,  where
$t$ is
defined $\nu_2\,=\,t\,\nu_1$. We have argued that in the  Regge approach
we expect that $\nu_2$ is proportional to $\nu_1$. One can see from
Fig.7, that $R_D$ is about 0.4, only for large values of $t$. Note
that $\nu_{1,2}\,\,=\,\,\sqrt{t}\,\nu_1$ at high energies.
Therefore, for $t\,>\,100$ we have $\nu_{1,2}\,>\, 10
\nu_1$ in accordance with  the global fit of the experimental data on ``soft"
processes\cite{GLMSFL}.

\begin{figure}
\begin{tabular} {c c}
\epsfig{file=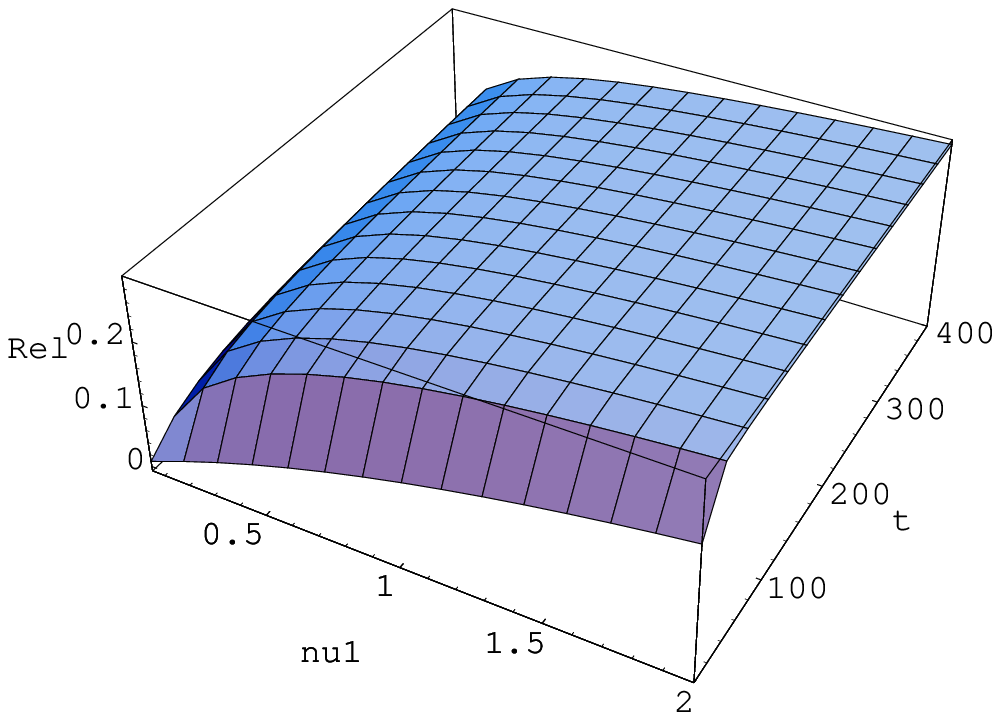,width= 70mm} & \epsfig{file=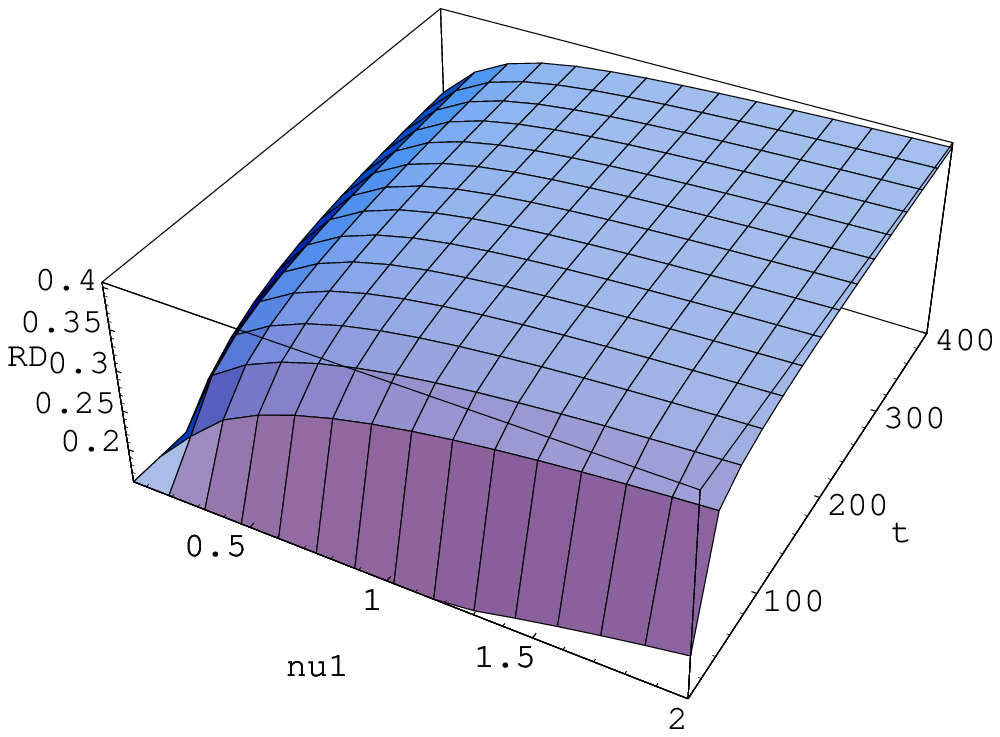,width=
70mm}\\
Fig. 7-a & Fig.7-b\\
\end{tabular}
\caption{\it Ratios $R_{el} \,=\,\s_{el}/\s_{tot}$ and $R_D\,=\,( \s_{el}
+ \s_{SD} + \s_{DD} )/\s_{tot} $ versus $\nu_1$ and $t$, where $t$ is
defined as $\nu_2\,=\,t\,\,\nu_1$.}
 \label{Fig.7}
\end{figure}

In Fig.8 we take $t$ = 300 and plot $R_D$ and $R_{el} $ versus $\nu_1$ .
One can see that $R_D$ is a very smooth function,, while
$R_{el}$ depends substantially on $\nu_1$. Such a behaviour reflects
the experimental situation shown in Fig.4. We use the experimental
data for $R_{el}$ given in Fig.4 to assign a definite value of
$\nu_1$ for  a definite value of energy. In particular, we find
$\nu_1
= 0.25 $ for $\sqrt{s}\, = \,640 \,GeV$ and $\nu_1 = 0.5 $ for
 $\sqrt{s}\,=\,1800\,GeV$ to give $R_{el} = 0.187 $ and $R_{el} = 0.237$,
 respectively.

Comparing Fig.8a and Fig.8b we can see that $\b$ - dependence for
$\b\,>\,0.5$ is not  essential.

\begin{figure}
\begin{tabular} {c c}
\epsfig{file=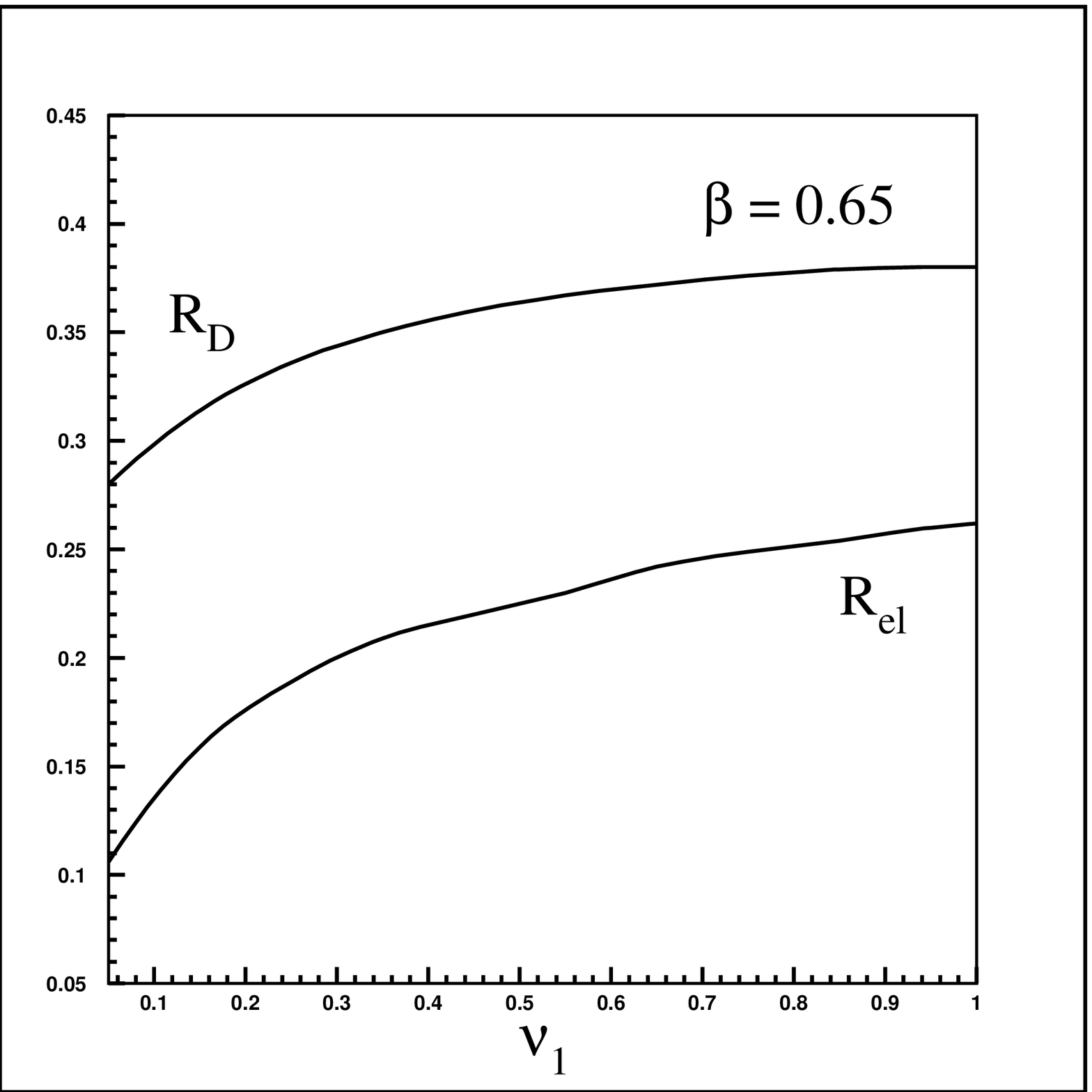,width= 80mm} &\epsfig{file=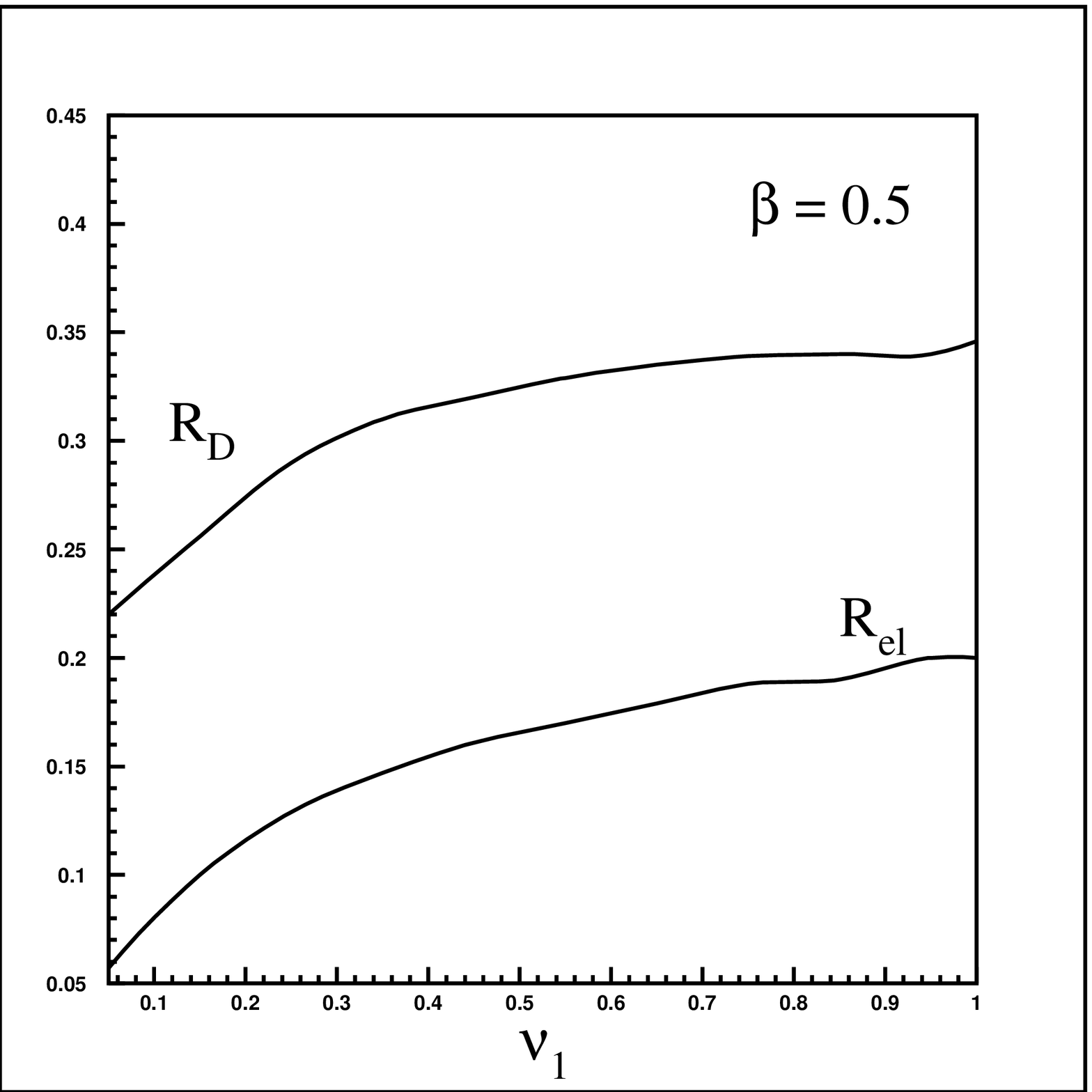,width= 80mm}\\
Fig. 8-a & Fig.8-b\\
\end{tabular}
\caption{\it Ratios $R_{el} \,=\,\s_{el}/\s_{tot}$ and $R_D\,=\,( \s_{el}
+ \s_{SD} + \s_{DD} )/\s_{tot} $ versus $\nu_1$ for two values of $\b$ (
$\b $ = 0.65 ( Fig.8a ) and $\b$ = 0.5 ( Fig.8b ).}
 \label{Fig.8}
\end{figure}

\subsection{``Hard" radii}

Before calculating the value of the survival probability we 
discuss the values of the ``hard" radii $\tilde{R}^2_{i,k}$.
Fortunately, these can be  determined   directly from the experimental
data on
diffractive production of vector mesons in DIS\cite{HERADDDATA} at
HERA. The data show that this production  depends  differently  on the
momentum transfer for elastic ( see, for example,  the process
$\gamma^* + p\,\ra\,\Psi + p $ in Fig. 9 ) and inelastic ( $\gamma^*
+ p\,\ra\,\Psi + X $ in Fig.9 ) production. In the  exponential
parameterization, the $t$ - dependence are  characterized by the
slopes $B_{el}$ and $B_{in}$.

Experimentally, these slopes are $B_{el}\,\,=\,\,4\,GeV^{-2} $ and
$B_{in}\,\,=\,\,1.66\,GeV^{-2}$. Since the vertex $\gamma^*
\,\ra\,\Psi$ does not depend on $t$ at large value of photon virtuality
$Q^2$,
we can view such an  experiment as a  way of measuring the
 $t$
- dependence of the Pomeron - hadron form factor, and/or  the
transition form factor of a hadron to a diffractive state, due to
Pomeron exchange. Incorporating  the experimental data on slopes in the
expressions of the ``hard"
radii we find
\bea \label{R1}
& \tilde{R}^2_{1,1}\,\,\,\,=\,\,\,\,16\,\,GeV^{-2}\,\,;&\nonumber\\
& \tilde{R}^2_{1,2}\,\,\,\,=\,\,\,\,11.32\,\,GeV^{-2}\,\,;&\\
&\tilde{R}^2_{2,2}\,\,\,\,=\,\,\,\,6.64\,\,GeV^{-2}\,\,.&\nonumber
\eea
One can see from \eq{S3}  the value of the survival
probability depends on the ratios $r_{i,k}$.  To calculate these ratios
we have to specify the values of ``soft" radii $R^2_{i,k}$. We assumed
the following  values for the ``soft" radii
\bea \label{R2}
&
R^2_{1,1}\,\,\,\,=\,\,\,\,12\,\,\,+\,\,4\,\alpha'_P(0)\,\,\ln(s/s_0)
\,\,GeV^{-2}\,\,;&\nonumber\\
& R^2_{1,2}\,\,\,\,=\,\,\,6\,\,\,+\,\,4\,\alpha'_P (0)\,\,\ln
(s/s_0)\,\,GeV^{-2}\,\,;&\\
&R^2_{2,2}\,\,\,\,=\,\,\,\,4\,\alpha'_P\,\ln(s/s_0)
\,\,GeV^{-2}\,\,.&\nonumber
\eea
It should be stressed that these radii are taken  from our
parameterization of the ``soft" data, but they also  describe the data on
the elastic slope, and agree with known information on the  $t$ -
dependence of  single diffraction dissociation\cite{DDDINO}.

\begin{figure}
\centerline{\epsfig{file=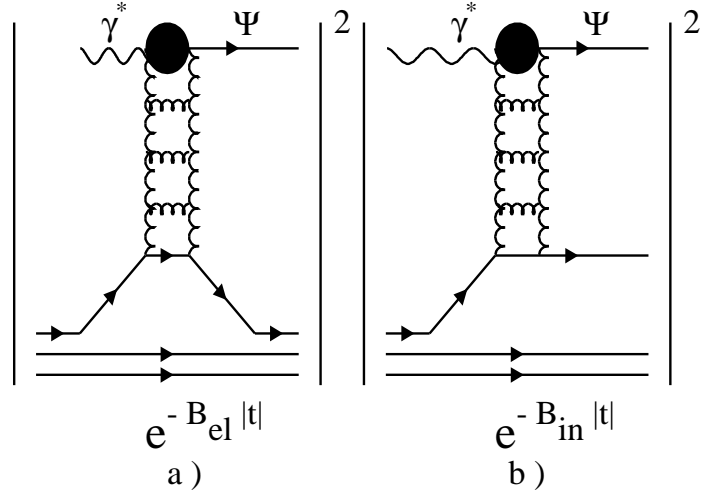,width=120mm}}
\caption{ \it Two slopes in diffractive J/$\Psi$ production in DIS in
the Additive Quark Model.}
 \label{Fig.9}
\end{figure}

\subsection{ $\mathbf{< \mid S^2_{spectator} (\,s\,) \mid >}$}

In Fig.10 ( solid curve ) we show  the value of the survival probability
at
$\sqrt{s} \,\,=\,\,1800\,GeV$ as a function of $\b$. This value
turns out to be rather small, in agreement with the experimental
data\cite{D0}.

Fig.11 ( solid curve )  shows the energy dependence of  $< \mid
S^2_{spectator}(\,s\,)\mid^2 >$, namely the ratio
\beq \label{R3}
R\,\,\,\,=\,\,\,\frac{< \mid S^2_{spectator}
(\,\sqrt{s}\,=\,640\,\,GeV\,)\mid^2>}{< \mid S^2_{spectator}
(\,\sqrt{s}\,=\,1800\,\,GeV\,)\mid^2 >}\,\,,
\eeq
versus $\b$.

In both these figures, at each fixed $\b$, we found the values of  $\nu_1$
for $\sqrt{s} =
640\,GeV$ and for  $\sqrt{s} =  1800\,GeV$ from the values of $R_{el}$
( $R_{el} \,=\,0.187 $ and $R_{el}\,=\,0.237$, respectively, see Fig.4a ).
The value of the survival probability was calculated using \eq{S3}.

One can see that the ratio $R_S$  reaches the value of 2.2
for $\b = 0.6$, which we consider  a typical value which fits the
experimental data. It should be stressed that for $\b\,<\,0.5$ the value
of $R_D$ is smaller than 0.3. This fact is in contradiction with the
experimental data shown in Fig.4b. At $\b = 0$  we 
reduce to the usual Eikonal
Model and we obtain a ratio $R_S$ which    is about 1.6. In
Ref.\cite{GLMSP2}   we  found  a larger value as we used  an average
value for the ``hard" radius, while here we have  introduced different
radii for the
different processes.

We would like to stress, that the  accuracy of our estimates is not
high. This is mostly because of  the large dispersion of the experimental
data
for $R_{el}$ ( see Fig.4a ). To illustrate this point we plot the value of
 $< \mid S^2 \mid >$  in Fig.10 ( dotted line ) taking  $R_{el}$ = 0.215
at $\sqrt{s} = 1800\,GeV$. One can see that the difference between these
two curve is about a factor of  2.  The spread of the
experimental data influences  dramatically the ratio of \eq{R3} ( see
Fig.11 ). To
illustrate it we plot in Fig.11 several lines that correspond to different
choices of $R_{el}$
\bea
& R_{el}(\sqrt{s}\,=\,640\,GeV )\,\,=\,\,0.187 \,\,\, and\,\,\,
R_{el}(\sqrt{s}\,=\,1800\,GeV )\,\,=\,\,0.237\,\,;&\label{C1}\\
& R_{el} ( \sqrt{s}\,=\,640\,GeV )\,\,=\,\,0.212 \,\,\, and
\,\,\,R_{el}(\sqrt{s}\,=\,1800\,GeV )\,\,=\,\,0.237 \,\,;&\label{C2}\\
& R_{el} ( \sqrt{s}\,=\,640\,GeV )\,\,=\,\,0.187\,\,\, and\,\,\,
R_{el}(\sqrt{s}\,=\,1800\,GeV )\,\,=\,\,0.215\,\,.&\label{C3}
\eea
The full line in Fig.11 corresponds to parameters of \eq{C1}, the dashed
line to parameters of \eq{C2} and the dash-dotted line  to parameters
of \eq{C3}.

We conclude, therefore,  that
  we cannot
obtain a definite prediction even within the framework of a
particular model, due to the  uncertainties in  the experimental
data on $R_{el}$.

\begin{figure}
\centerline{\epsfig{file=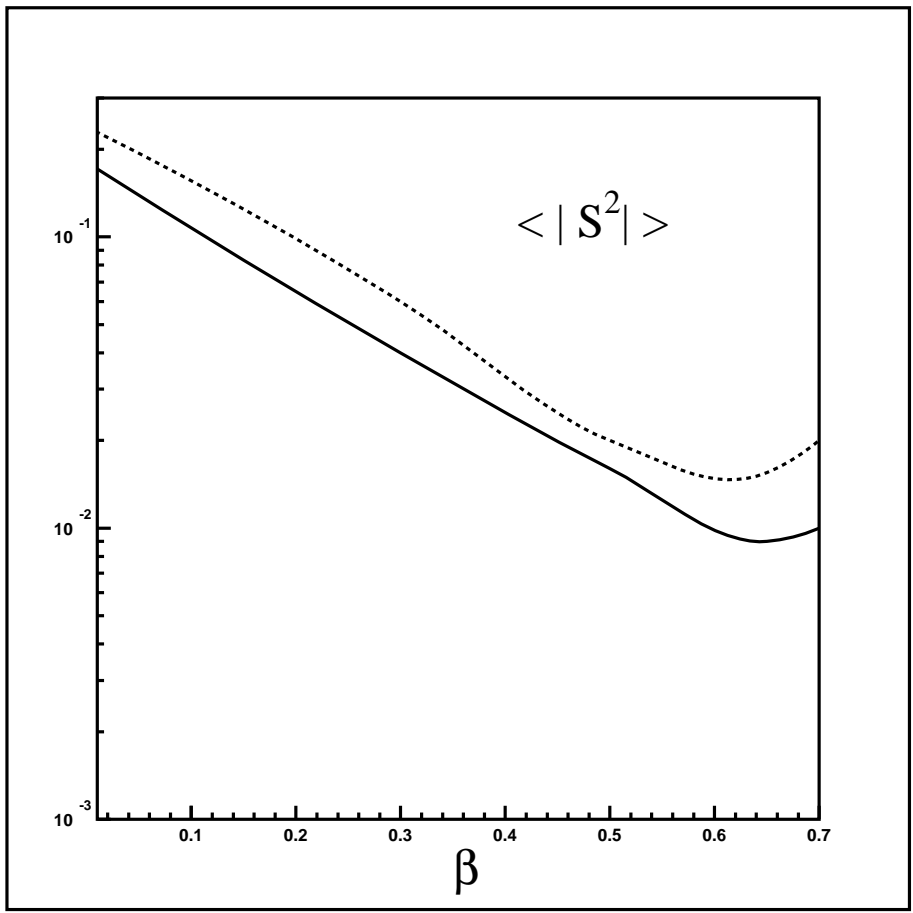,width=120mm}}
\caption{ \it The value of $< \mid S^2 \mid >$ at $\sqrt{s} = W = 1800
GeV$ versus $\b$ in the three channel  model. The full and dashed lines
corresponds to parameters of \eq{C1} and \eq{C2}, respectively.}
 \label{Fig.10}
\end{figure}

\begin{figure}
\centerline{\epsfig{file=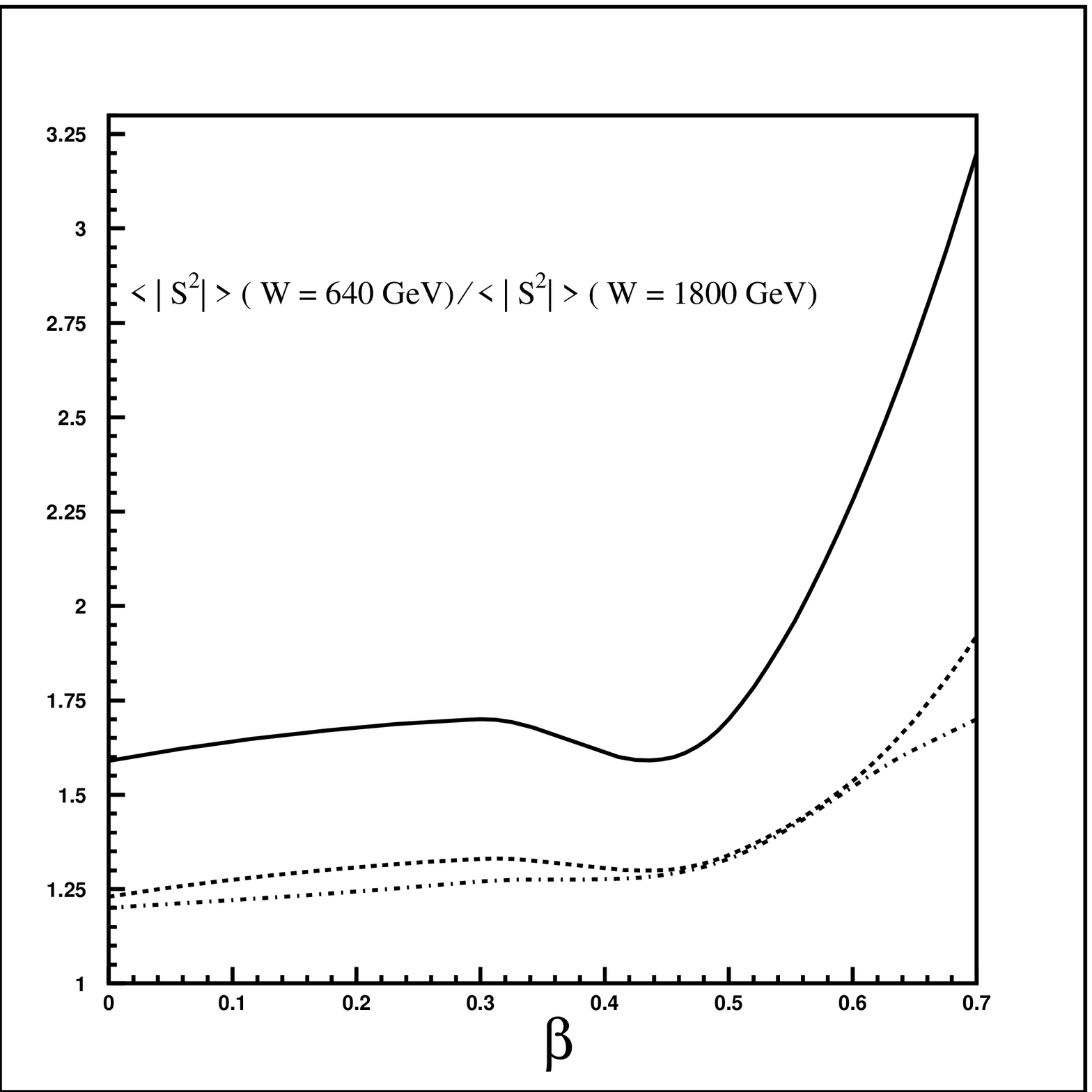,width=120mm}}
\caption{ \it The value of ratio $< \mid S^2 \mid > ( W = 640\,GeV )/
< \mid S^2 \mid > ( W = 1800\,GeV )$
  versus $\b$ in the three channel model. W = $\sqrt{s}$.The full,
dashed  and dash-dotted lines
corresponds to parameters of \eq{C1}, \eq{C2} and \eq{C3}, respectively }
 \label{Fig.11}
\end{figure}

\section{Discussion and Summary}

In this paper we give an example of a model in which
\begin{enumerate}

\item\,\,\, The processes of elastic and  diffractive rescatterings have
been
taken into
account, their
cross
sections being of the same order of magnitude  ( $\s_{SD} \,\approx
\s_{el} $ and
$\s_{DD}\,\approx\,\s_{el}$ )\,\,;

\item \,\,\, It was shown that the scale of the SC is not  given  by
the
ratio $R_D$, but rather by the separate  ratios $ \s_{el}/\s_{tot}$,
$\s_{SD}/\s_{tot}
$
and $\s_{DD}/\s_{tot}$. Since each of these ratios shows
considerable energy dependence,  we do not expect a constant survival
probability,  contrary to  the simpler model of Ref.\cite{RR};

\item\,\,\,  It was demonstrated that the small value of the survival
probability, as well as its strong energy dependence  appear
naturally in our approach;

\item\,\,\, The  rather large value of $\nu_2 \,\approx \,\,300 \,\nu_1$
reflects (i) a smaller value  of $R^2_{2,2}$ in comparison with
$R^2_{1,1}$
observed experimentally, and (ii) the fact that this value takes into
account the integration over the mass of the produced hadrons in our
oversimplified model;

\item\,\,\, The parameters that have been used are in agreement with the
more
detailed  fit of the experimental data on ``soft" processes 
( see Ref.\cite{GLMSFL} ).

 \end{enumerate}

Theoretical predictions for the
value of the survival probability are still not very reliable. However,
developing different models enables us to learn and assess which class of
models provides natural predictions for both the value and the
energy behaviour of the survival probability. We want to draw the readers
attention to the fact that our estimates for  the value of the survival
probability given in Figs. 10 and 11 are  very close to the estimates
obtained  in
the
Eikonal Model\cite{GLMSP2}, in spite of the fact that the three channel
model is quite different from the eikonal one.

New    measurements both on  LRG processes, and on the
cross sections of  diffraction dissociation ( in particular, on
double diffraction ) would be very useful for  a deeper
understanding  of ``soft"
interactions at high energy.

{\bf Acknowledgements:} This research was supported in part by the Israel
Science Foundation, founded by Israel Academy of Science and Humanities.

\end{document}